\documentclass{article}
\usepackage{graphicx}
\usepackage{xcolor}
\usepackage{amsmath, amsthm, amssymb}
\usepackage{times}

\newcommand{\bmp}{\begin{minipage}}
\newcommand{\emp}{\end{minipage}}

\newcommand{\Rt}{\mbox{\it R}^{\top}}
\newcommand{\Rb}{\mbox{\it R}^{\bot}}
\newcommand{\Ra}{\mbox{\it Set}_R}
\newcommand{\SL}{\mbox{\it SL}}
\newcommand{\SR}{\mbox{\it SR}}
\newcommand{\PopL}{\mbox{\it Pop}_L}
\newcommand{\PopR}{\mbox{\it Pop}_R}
\newcommand{\PushLR}{\mbox{\it Push}_{LR}}
\newcommand{\Id}{\mbox{\it Id}}
\newcommand{\tp}{\mbox{\it top}}
\newcommand{\LR}{\mbox{\it LR}}
\newcommand{\M}{\mbox{\it MinMax}}

\newcommand{\LpRp}{L^+R^+}
\newcommand{\LpRm}{L^+R^-}
\newcommand{\LmRm}{L^-R^-}
\newcommand{\LmRp}{L^-R^+}
\newcommand{\FindL}{\mbox{\it Find}_L}
\newcommand{\mint}{\mbox{\it m}_t}
\newcommand{\maxt}{\mbox{\it M}_t}
\newcommand{\Pm}{P^{-1}}
\newcommand{\BL}{\mbox{\it b}}
\newcommand{\BR}{\mbox{\it B}}
\newcommand{\n}{[n]}
\newcommand{\bt}{\mbox{\it b}_t}
\newcommand{\Bt}{\mbox{\it B}_t}
\newcommand{\out}{{Filter}}
\newcommand{\up}{\mbox{\it Inf}}
\newcommand{\st}{\mbox{\it Pairs}_t}
\newcommand{\mini}{\mbox{\it min}}
\newcommand{\maxi}{\mbox{\it max}}
\newcommand{\ISP}{\mbox{ISP}}
\newcommand{\C}{\mathcal{C}}
\newcommand{\PK}{\mathcal{P}}

\newcommand{\Inf}{\mbox{\it Small\,}}
\newcommand{\fort}{{\bf for}$_t$}

\newcommand{\next}{\mbox{\it next}}
\newcommand{\nextu}{\mbox{\it nextt}}
\newcommand{\query}{\mbox{\it query}}

\newcommand{\Position}{\mbox{\it Position}}
\newcommand{\sign}{\mbox{\it sign}}
\newcommand{\Fir}{\mbox{\it First}}

\newtheorem{thm}{Theorem}
\newtheorem{cor}{Corollary}
\newtheorem{fait}{Claim}
\newtheorem{rmk}{Remark}
\newtheorem{defin}{Definition}
\newtheorem{ex}{Example}

\newcommand{\br}{\begin{rmk}\rm}
\newcommand{\er}{\end{rmk}}
\newcommand{\bdefin}{\begin{defin}\rm}
\newcommand{\edefin}{\end{defin} }
\newcommand{\bex}{\begin{ex}\rm}
\newcommand{\eex}{\end{ex}}

\newcommand{\bthm}{\begin{thm}}
\newcommand{\ethm}{\end{thm}}
\newcommand{\bcor}{\begin{cor}}
\newcommand{\ecor}{\end{cor}}
\newcommand{\bfn}{\begin{fait}}
\newcommand{\efn}{\end{fait}}

\usepackage{algorithm}
\usepackage{algorithmic}

\makeatletter
\newcommand{\setalglineno}[1]{%
  \setcounter{ALC@line}{\numexpr#1-1}}
\makeatother

\renewcommand{\Box}{\rule{1.5mm}{3mm}}

\begin{document}


\begin{center}
{\bf\large $\M$-Profiles: A Unifying View of Common Intervals, Nested Common\\  
\vspace*{0.2cm}

Intervals and Conserved Intervals of $K$ Permutations}\\

%

\vspace*{1cm}

Irena Rusu\footnote{Irena.Rusu@univ-nantes.fr}

L.I.N.A., UMR 6241, Universit\'e de Nantes, 2 rue de
la Houssini\` ere,\\

 BP 92208, 44322 Nantes, France
\end{center}


\vspace*{1cm}

\hrule
\vspace{0.3cm}

\noindent{\bf Abstract} 

Common intervals of $K$ permutations over the same set of $n$ elements were firstly investigated
by T.~Uno and M.Yagiura (Algorithmica, 26:290:309, 2000), who proposed an efficient algorithm
to find common intervals when $K=2$. Several particular classes of intervals 
have been defined since then, {\it e.g.} conserved intervals and nested common intervals,
with applications mainly in genome comparison. Each such class, including common intervals,
led to the development of a specific algorithmic approach for $K=2$, and - except for 
nested common intervals - for its extension to an arbitrary $K$. 

In this paper, we propose a common and efficient algorithmic framework for finding
different types of common intervals in a set $\PK$ of $K$ permutations, with arbitrary $K$. 
Our generic algorithm is based on a global representation of the information stored in $\PK$, called 
the $\M$-profile of $P$, and  an efficient data structure, called an $\LR$-stack, 
that we introduce here.
We show that common intervals (and their subclasses of irreducible common intervals 
and same-sign common intervals), nested common intervals (and their subclass of maximal nested common intervals) as
well as conserved intervals (and their subclass of irreducible conserved intervals)
may be obtained by appropriately setting the parameters of our algorithm in each case.
All the resulting algorithms run in $O(Kn+N)$-time and need $O(n)$ additional space,
where $N$ is the number of solutions.
The algorithms for nested common intervals and maximal nested common intervals are
new for $K>2$, in the sense that no other algorithm has been given so far to solve the problem
with the same complexity, or better. The other algorithms are as efficient as the best known 
algorithms.
\bigskip

\vspace{0.2cm}

\hrule

\section{Introduction}
Common, conserved  and nested common intervals of two or more permutations have been defined  
and studied in the context of genome comparison. Under the assumption that a set of genes
occurring in neighboring locations within several genomes represent functionally related genes 
\cite{galperin2000s,lathe2000gene,tamames2001evolution}, common intervals and their subclasses are used 
to represent such conserved regions, thus helping for instance to detect 
clusters of functionally  related genes \cite{overbeek1999use, tamames1997conserved}, 
to compute similarity  measures between genomes \cite{BergeronSim, AngibaudHow} 
or to predict protein functions \cite{huynen2000predicting, von2003string}. 
Further motivations and details may be found in the papers introducing these intervals,
that we cite below. 

In these applications, genomes may be represented either as permutations, when they do not
contain duplicated genes, or as sequences. In sequences, duplicated genes usually play
similar roles  and lead to a more complex interval search \cite{didier2003common, schmidt2004quadratic},  but sometimes they are appropriately matched 
and renumbered so as to obtain 
permutations  \cite{blin2005conserved, AngibaudApprox}. 

We focus here on the case of permutations.
Efficient algorithms exist for finding common and conserved intervals in $K$ permutations 
($K\geq 2$), as well as for finding irreducible common
and irreducible conserved intervals \cite{UnoYagura,heber2011common, BergeronK, BergeronSim}. Nested common intervals and maximal nested common intervals
have been studied more recently \cite{hoberman2005incompatible}, and efficient algorithms exist only for the case of two permutations \cite{blin2010finding}.

Surprisingly enough, whereas all these classes are subclasses of common intervals, 
each of them has generated a different analysis, and a different approach to obtain
search algorithms. Among these approaches, interval generators \cite{BergeronK} have
been shown to extend from common intervals to conserved intervals \cite{rusu2011new}, but this
extension is not easily generalizable to other subclasses of common intervals.

The approach we present in this paper exploits the natural idea that an efficient
algorithm for common intervals should possibly be turned into an efficient algorithm for 
a subclass of common intervals by conveniently setting some parameters
so as to filter the members of the subclass among all common intervals. It also chooses
a different viewpoint with respect to the information to be considered. Instead
of searching intervals directly in the permutations, it first extracts the helpful
information from the permutations, focusing on each pair $(t,t+1)$ of successive values
in $\{1, 2, \ldots, n\}$ and defining the so-called $\M$-profile of the permutations. 
Then, it progressively computes the set of interval candidates,
but outputs them only after a filtering procedure selects the suitable ones.

The organization of the paper is as follows. In Section \ref{sect:generalities}, we present
the main definitions and the problem statement. Then, in Section~\ref{sect:main}, we introduce the abstract data structure on which strongly relies our main algorithm called $\LR$-Search,
also described in this section. 
The complexity issues are discussed in Section \ref{sect:complex}. In Section~\ref{sect:comm} we give 
the specific settings  of our algorithm for  common, nested 
common and conserved  intervals, and prove the correctness in each case. 
In Section~\ref{sect:next} we show how to further modify the algorithm so as to deal with even
smaller subclasses. Section \ref{sect:concl} is the conclusion.

\section{Generalities}\label{sect:generalities}

For each positive integer $u$, let  $[u] := \{1, 2, \ldots, u\}$. Let $\PK :=\{P_1, P_2, \ldots, P_K\}$ be a set of permutations 
over $\n$, with $n>0$ and integer.  
The {\em interval}  $[i,j]$ of  $P_k$, defined only for 
$1\leq i< j\leq n$, is the set of elements located between position $i$ (included) and position $j$ (included) in $P_k$.
Such an interval is denoted $(i..j)$ 
when $P_k$ is the identity permutation $\Id_n : = (1\, 2\, \ldots\, n)$.  Then $(i..j)=\{i,$ $i+1, 
\ldots, j\}$. For an interval $[i,j]$ of $P_k$, we also say that the interval is {\it delimited} 
by its elements located at positions $i$ and $j$, 
 or equivalently that  
these elements are the {\it delimiters} of $[i,j]$ on $P_k$ (note the difference between delimiters, which
are {\it elements},  and their {\it positions}). Furthermore, let $\Pm_k:\n\rightarrow\n$ be the function, 
easily computable in linear time, 
that associates with every element of $P_k$ its position in $P_k$. We now define common intervals:

\bdefin   \cite{UnoYagura}
A {\em common interval} of $\PK$ is a set of integers that is an interval of each $P_k$, $k\in[K]$.
\edefin 

Nested common intervals are then defined as follows:

\bdefin  \cite{hoberman2005incompatible}
A {\em nested common interval} (or {\em nested interval} for short) of $\PK$ 
is a common interval $I$ of $\PK$ that either satisfies $|I|= 2$, or contains 
a nested interval of cardinality $|I|-1$.  
\edefin 

\bex
Let $P_1=\Id_7$ and $P_2=(7\, 2\, 1\, 3\, 6\, 4\, 5)$. Then the common intervals of 
$\PK=\{P_1, P_2\}$ are $(1..2), (1..3), (1..6), (1..7), (3..6), (4..5)$ and
$(4..6)$, whereas its nested intervals are $(1..2), (1..3), (3..6)$, $(4..5)$ and $(4..6)$.
\label{ex:commnest}
\eex

With the aim of introducing conserved intervals, define now a {\it signed} permutation
as a permutation $P$ associated with a boolean vector $\sign_P$ that provides a $+$ or $-$ sign
for every element of $P$. Then $\sign_P[i]$ is the sign of the integer $i$ in $P$. 
An element of $P$ is called {\em positive} or {\em negative} if its associated
sign is respectively $+$ or $-$.  A permutation is then a signed permutation containing only 
positive elements.
 
\begin{rmk} Note that, even in a signed permutation, the elements are {\it positive
integers}. This assumption greatly simplifies our algorithms. However, for a better 
understanding of our examples, we use the notation $x$ or $+x$ for a positive element,
and $-x$ for a negative element. The reader should however notice the precise definition
of the vector $\sign_P$, which will be used in a proof in Section \ref{subsect:conserved}.
\end{rmk}

According to \cite{BergeronSim}, we give now the definition of a conserved interval.

\bdefin   \cite{BergeronSim}
Let $\PK :=\{P_1, P_2, \ldots, P_K\}$ be a set of signed permutations over $\n$, such that the
first element of $P_k$ is $1$ (positive) and the last element of $P_k$ is $n$ (positive), for each $k\in[K]$.  
A {\em conserved interval} of $\PK$ is a 
common interval of $\PK$ (ignoring the signs) delimited on  $P_k$ by the values
$a_k$ (left) and $b_k$ (right), for each $k\in[K]$, such that exactly one of the following 
conditions holds for each $k$:

(1) $a_k=a_1$ with the same sign and $b_k=b_1$ with the same sign. 

(2) $a_k=b_1$ with different signs and $b_k= a_1$ with different signs. 

\edefin  

\bex
Let $P_1=\Id_7$ and $P_2=(1\, \mbox{-3}\, \mbox{-2}\,\, 6\, \mbox{-4}\, \mbox{-5}\,\, 7)$ (equivalently, we have $\sign_{P_2}=[+, -, -, -, -,$ $ +, +]$).
Then the conserved intervals of 
$\PK=\{P_1, P_2\}$ are $(1..7)$ and  $(2 ..3)$, whereas its common intervals (ignoring signs)
are $(1..3), (1..6), (1..7), (2..3), (2..6), (2..7), (4..5),  (4..6), (4..7)$.
\label{ex:cv}
\eex

In the following problem, $\C$ refers to a class of common intervals, {\it e.g.}  common, nested or conserved.
\smallskip

\noindent{$\C$-{\sc Interval Searching Problem} (abbreviated $\C$-\ISP)\\
\noindent\begin{tabular}{ll}
 
\hspace*{-0.2cm}{\bf Input:} & \begin{minipage}[t]{12.5cm} A set  $\PK=\{P_1, P_2, \ldots, P_K\}$ of signed permutations over $\n$,
satisfying the conditions required by the definition of $\C$.\smallskip\end{minipage}\\
\hspace*{-0.2cm}{\bf Requires:}& \hspace*{-0.05cm}Give an efficient algorithm to output without redundancy all $\C$-intervals of $\PK$. 

\end{tabular}
\bigskip

In all cases, we may assume without loss of generality that $P_1=\Id_n$, by
appropriately renumbering $P_1$. The other permutations must be renumbered accordingly.

In this paper, we propose a common efficient algorithm to solve $\C$-\ISP\, when $\C$ stands for common, 
nested, conserved intervals as well as for their respective subclasses of 
irreducible common, same-sign common, maximal nested and irreducible conserved intervals (defined in
Section~\ref{sect:next}).   For 
common and conserved intervals (irreducible or not), efficient algorithms have been proposed 
so far, developping different and sometimes very  complex approaches. 
For nested and maximal nested intervals, efficient algorithms exist for the case $K=2$.
We solve here the case of an arbitrary $K$. Our approach is common for all classes, 
up to the filtering of the intervals in $\C$ among
all common intervals. 

\section{Main Algorithm}\label{sect:main}

Our $\LR$-Search algorithm is based on two main ideas. First,
it gathers information from $\PK$ that it stores as {\it anonymous} constraints on each pair 
$(t,t+1)$ of successive values, since during the algorithm it is useless to know the
source of each constraint. Second, it fills in a data structure that allows us
to find {\it all} common intervals with provided constraints if we need, but an additional \out\, 
procedure is called to choose and to output only the intervals in the precise class $\C$
for which the algorithm is designed.

The algorithm uses a specific data structure that
we call an $\LR$-stack. The candidates for the left (respectively right) endpoint of a common 
interval are stored in the $L$ (respectively $R$) part of the $\LR$-stack. At each step of the
algorithm, the $\LR$-stack is updated, the solutions just found are output, and the $\LR$-stack is
passed down to the next step.

\subsection{The $\LR$-stack}

The $\LR$-stack is an abstract data structure, whose efficient implementation is discussed in Section \ref{sect:complex}.

\bdefin  An {\em $\LR$-stack} for an ordered set $\Sigma$ is a 5-tuple $(L, R, \SL, \SR, \Rt)$ such that:

\begin{itemize}
\item $L, R$ are stacks, each of them containing distinct elements from $\Sigma$ in either increasing or 
decreasing order (from top to bottom). The {\it first} element of a stack is its top, the {\it last} one is its bottom.
\item $\SL,\SR \subset \Sigma$ respectively represent the set of elements on $L$ and $R$.
\item $\Rt : \SL\rightarrow \SR$ is an injection that associates with each $a$ from $\SL$ a pointer to an 
element on $R$ such that $\Rt(a)$ is before $\Rt(a')$ on $R$ iff $a$ is before $a'$ on $L$.
\end{itemize}

According to the increasing (notation +) or decreasing (notation -) order of the elements on
$L$ and $R$ from top to bottom, an $\LR$-stack may be of one of the four types $\LpRp,  \LmRm, \LpRm, \LmRp$.
\edefin 

\br We assume that each of the stacks $L,R$ admits the classical operations 
$pop, push$, and that their elements may be read without removing them. 
In particular, the function $\tp()$ returns the first element of the stack, without removing it,
and the function $\next(u)$ returns the element immediately following $u$ on the stack
containing $u$, if such an element exists. 
\label{rmk:next}
\er

  We further denote, for each $a\in\SL$ and with $a'=\next(a)$, assuming that $\next(a)$ exists:
$$\Ra(a)=\{b\in\SR\, |\, b\,\,  \mbox{\rm is located on}\,  R\,\, \mbox{\rm between}\, \Rt(a)\, \mbox{\rm included and}\,  \Rt(a')\, \mbox{\rm excluded}\}$$
When $\next(a)$ does not exist, $\Ra(a)$ contains all elements between $\Rt(a)$ included and the bottom of $R$ included.
Then $\Rt(a)$ is the first (i.e. closest to the top) element of $\Ra(a)$ on $R$.

The following operations on the LR-stack  are particularly useful.
Note that they do not affect the properties of an $\LR$-stack. Sets $\Ra()$ are assumed to be updated without further specification whenever the pointers $\Rt()$ change.
Say that $a'$ is {\it $L$-blocking} for $a$, with $a'\neq a$, if $a$ cannot be pushed on $L$ when $a'$ is
already on $L$ (because of the increasing/decreasing order of elements on $L$),
and similarly for $R$.

\begin{itemize}
\item $\PopL(a)$, for  some $a\in \Sigma$: 
pop successively from $L$ all elements that are $L$-blocking for $a$, push $a$ on $L$
iff at least one $L$-blocking element has been found and $a$ is not already on $L$, 
and define $\Rt(\tp(L))$ as $\tp(R)$. At the end,
either $a$ is not on $L$ and no $L$-blocking element exists for $a$, or $a$ is on the top of $L$ and $\Rt(a)$ is a pointer to the top of $R$.

\item $\PopR(b)$, for some $b\in\Sigma$: pop successively from $R$ all elements that
are $R$-blocking for $b$, update all pointers $\Rt()$ and successively pop from $L$ all the elements $a$
with $\Rt(a)=nil$.  At the end, either $b$ is not on $R$ and no $R$-blocking element exists for $b$, or $b$ is on the top of $R$.

\item $\PushLR(a,b)$, for some $a,b\in \Sigma$ (performed when no $L$-blocking element exists for $a$ and
no $R$-blocking element exists for $b$): push $a$ on $L$ iff $a$ is not already on the top of $L$, 
push $b$ on $R$ iff $b$ is not already on the top of $R$,
and let $\Rt(\tp(L))$ be defined as $\tp(R)$.

\item $\FindL(b)$, for some $b\in\Sigma$: return the element $a$ of $\SL$ such that $b\in \Ra(a)$.

\end{itemize}

\bex
In Figure \ref{fig:LRSearch}, consider the $\LmRp$-stack below the pair $(4,5)$. Here, $\Ra(4)=\{5,6\}$
and $\Ra(1)=\{7\}$. The instruction $\PopL(3)$ discards $4$ from $L$, and pushes $3$ instead, also
defining $\Rt(3)=5$. Next, the instruction $\PopR(6)$ discards $5$ from $R$. The resulting $\LR$-stack
is represented below the pair $(3,4)$.
\eex

\begin{rmk}
 Note that operations $\PopL(a)$ and $\PopR(b)$ perfom {\it similar} but not {\it identical} modifications
 on stacks $L$ and $R$ respectively. Indeed, $\PopL(a)$ pushes $a$ on $L$ if at least one element of $L$
 has been discarded and $a$ is not already on $L$, whereas $\PopR$ discards elements, but never pushes $b$ on $R$.
\end{rmk}

\begin{figure}[t]
\vspace*{-1cm}
\begin{center}
\hspace*{-1cm}\includegraphics[width=17cm]{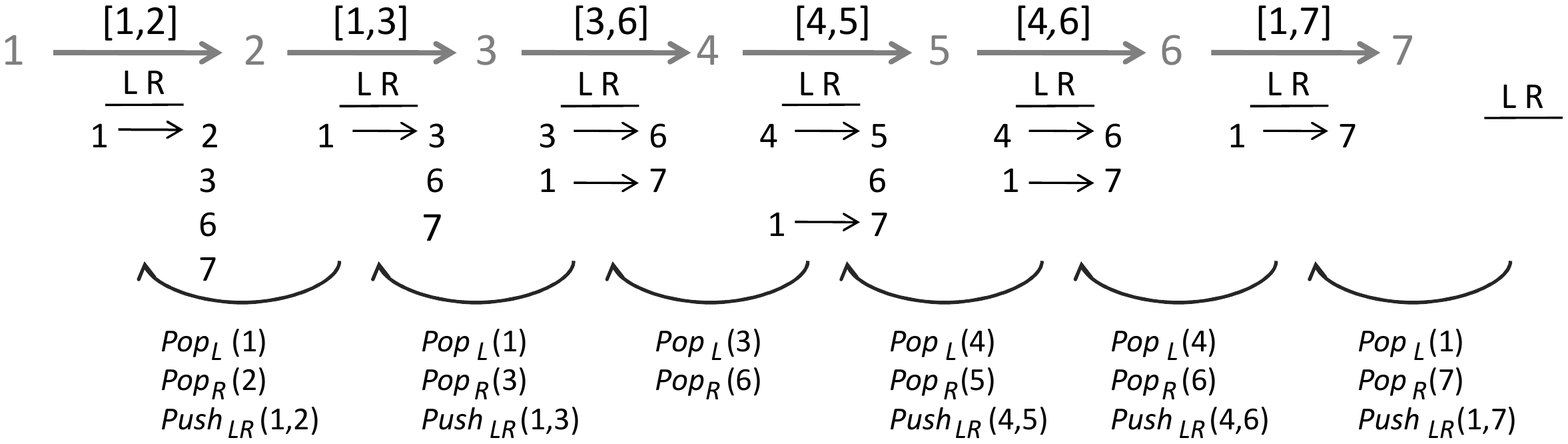}
\end{center}
\vspace*{-6.5cm}
\caption{$\M$-profile of $\PK=\{\Id_7, (7\, 2\, 1\, 3\, 6\, 4\, 5)\}$ with bounding functions
$\BL(t)=\mint$ and $\BR(t)=\maxt$, and execution of the $\LR$-Search algorithm.
The stack is initially empty. For each pair $(t,t+1)$ with $t= n-1, n-2, \ldots, 1$, the corresponding $\LR$-stack (below the pair) is 
obtained from the preceding $\LR$-stack (on its right) by performing the sequence of operations written
below the arrow linking the two $\LR$-stacks.}
\label{fig:LRSearch}
\end{figure}

\subsection{The $\LR$-Search algorithm}

Let $\PK$ be a set of $K$ signed permutations over $\n$. Recall that $P_1=\Id_n$. 
Now, let $\mint^k$ (respectively $\maxt^{\,k}$) 
be the minimum (respectively maximum) value in the interval of $P_k$ delimited by $t$ and $t+1$ (both included).
Also define 
$$\mint := \min\{\mint^k\,|\, 2\leq k\leq K\}, \maxt := \max\{\maxt^{\,k}\,|\, 2\leq k\leq K\}.$$
Notice that, for each $k\in[K]$, $\mint\leq \mint^k\leq t < t+1\leq \maxt^{\,k}\leq \maxt$.

Let the {\it bounding} functions $\BL,\BR:\n\rightarrow \n$ be two functions such that $\BL(t)\leq~\mint$ and $\BR(t)\geq \maxt$.  Denote $\bt :=\BL(t)$ and $\Bt :=\BR(t)$ for all $t\in [n-1]$. 

%
%
%
%
%

\bdefin
Let $\PK$ be a set of permutations on $\n$. Then the {\em $\M$-profile} of $\PK$ with respect to $\BL$ and $\BR$ 
is the set of pairs $[\bt,\Bt]$, $t\in[n-1]$. 
\edefin

The $\LR$-Search algorithm (See Algorithm \ref{algo:LR}) works as follows.   Each pair $(t,t+1)$, $1\leq t\leq n-1$, 
is associated with the couple $[\bt,\Bt]$, which means intuitively that every common interval $(a..c)$ of 
$\PK$ that contains $t$ and $t+1$ should satisfy $a\leq \bt < \Bt\leq  c$.
The $\LmRp$-stack,  initially empty, stores on $L$ (respectively on $R$) the candidates for the left
endpoint $a$ (respectively right endpoint $c$) of a common interval $(a..c)$, and links together the left and 
right candidates to form suitable pairs. Intuitively, a suitable pair is a pair of endpoints
for which no disagreement has been found yet. In the $\LR$-stack, a 
suitable pair is a pair $(a,c)\in \SL\times\SR$ such that $a\leq \FindL(c)$. Equivalently,
right candidate $c$ could form common intervals with $\FindL(c)$ as well as with all left 
candidates following $\FindL(c)$ on $L$.

\br
It is important to notice that, by definition, in an $\LmRp$-stack,  $a'$ is 
$L$-blocking for $a$ iff $a'>a$, and $b'$ is $R$-blocking for $b$ iff
$b'<b$.
\er

In its most general form, the algorithm outputs all common intervals without any
redundancy. However, in order to insure its efficiency even for strict subclasses $\C$ of common
intervals, it has two types of parameters allowing us to filter intervals during
the search (bounding functions achieve that) and during the output step (\out$\,\,$procedure
achieves that).  Then, $\LR$-Search considers (step 4) every pair $(t,t+1)$ using decreasing 
values of $t$, so as to output 
the intervals with same left endpoint altogether in the same step (increasing values of $t$
would produce the intervals with same right endpoint in the same step). Operations
$\PopL(\bt),\PopR(\Bt)$ (steps 5, 6) successively allow us to discard left candidates $a$  
that do not satisfy  $a\leq\bt$ (respectively right candidates $c$ that do not satisfy $\Bt\leq c$), 
and to update the suitable pairs accordingly. Steps 7-9 identify the cases where $t+1$ may be added to $R$,
namely when $\Bt=t+1$ ($\PushLR$ tests whether $t+1$ is already on $R$). Finally, in step 10  suitable 
intervals $(t..x)$  are necessarily common intervals since further constraints 
on pairs $(t',t'+1)$, $t'<t$, will not affect $(t..x)$. The \out$\,\,$procedure is then called 
to filter among all common intervals found here those that belong to some specific subclass.

\bex
Figure \ref{fig:LRSearch} shows the application of the $\LR$-Search algorithm on the
set $\PK=\{P_1,P_2\}$ of permutations, where $P_1=\Id_7$ and $P_2=(7\, 2\, 1\, 3\, 6\, 4\, 5)$.
In this case, we defined $\bt=\mint=\mint^2$ and $\Bt=\maxt=\maxt^2$. The constraints 
$[\bt,\Bt]$ are written above the arrow representing the pair $(t,t+1)$, for all $t\leq n-1$,
so that the upper part of the figure represents the $\M$-profile of $\PK$. 
The first  step corresponds to $t=6$, and consists in performing operations $\PopL(1)$,
$\PopR(7)$ and $\PushLR(1,7)$ on the initially empty stack. Thus the first two operations
have no effect, whereas the third one pushes 1 on $L$, $7$ on $R$ and defines $\Rt(1)$
as a pointer to 7. The next step takes the current state of the
$\LR$-stack and performs $\PopL(4), \PopR(6)$ and $\PushLR(4,6)$ to obtain the 
$\LR$-stack below the pair $(5,6)$. The first common intervals are output 
when $t=4$, namely $(4..5), (4..6)$.

\eex

\begin{algorithm}[t,boxed]
\caption{The $\LR$-Search algorithm}
\begin{algorithmic}[1]
\REQUIRE Set $\PK$ of signed permutations over $\n$, bounding functions $\BL$ and $\BR$,  \out$\,$ procedure
\ENSURE  All common intervals $(t..x)$ of $\PK$ with $x\in\Ra(t)$, filtered by \out

\STATE Compute $\mint^k,\mint$ and $\bt$ with $2\leq k\leq K$ and $t\in[n-1]$
\STATE Compute $\maxt^k,\maxt$ and $\Bt$ with $2\leq k\leq K$ and $t\in[n-1]$
\STATE Initialize an $\LmRp$-stack with empty stacks $L, R$
\FOR{$t$ from $n-1$ to $1$}
\STATE $\PopL(\bt)$ \hfill{\sl // discard from $L$ all candidates larger than $\bt$ and push $\bt$ instead}
\STATE $\PopR(\Bt)$ \hfill{\sl // discard from $R$ all candidates  smaller than $\Bt$}
\IF {$\Bt=t+1$}
\STATE $\PushLR(\bt,t+1)$ \hfill{\sl // $t+1$ is a new right candidate, suitable for each $a$ on $L$}
\ENDIF
\STATE Call \out$\,$ to choose a subset of intervals $(t..x)$ with $x\in \Ra(t)$
\ENDFOR
\end{algorithmic}
\label{algo:LR}
\end{algorithm}

\br
In the rest of the paper, the notation \fort\, concerns the execution of the {\bf for} loop in
$\LR$-Search for some fixed $t$. Similar notations will be used for the loops in the \out\, procedures given 
subsequently. When these notations are not confusing, we use them without any further specification.
\er

Let $\Ra^t(a)$ be the value of $\Ra(a)$ at the end of step 9 in \fort, for each $a$ on $L$. 
Let $\st$ be the set of pairs $(t..x)$ with $x\in \Ra^t(t)$.

\bex
In Figure \ref{fig:LRSearch}, $\Ra^4(4)=\{ 5,6\}$ and $\Ra^4(1)=\{ 7\}$, thus  ${\it Pairs}_4=\{(4..5),(4..6)\}$.
\eex

\bthm Assuming the \out$\,\,$procedure does not change the state of the $\LR$-stack, the 
set $A$ defined as $A~:~=\cup_{1\leq t<n}\st$ computed by $\LR$-Search is the set of all common intervals $(t..x)$ of $\PK$
(ignoring the signs) satisfying  $$t=\bt=\min\{b_w\,|\, t\leq w\leq x-1\}$$ 
$$x=B_{x-1}=\max\{B_w\,|\, t\leq w\leq x-1\}.$$

\label{thm:B}

\ethm

We first prove the following result. Notations
$\mini(I)$ and $\maxi(I)$ for an interval $I$ of $P$ respectively denote the minimum 
and maximum value in $I$.

\begin{fait}
Let $t$, $1\leq t<n$. After the execution of step 9 in \fort, 
we have $x\in \Ra^t(a)$ iff $t, x$ and $a$ satisfy the three conditions below:

\begin{enumerate}
\item[$(a)$] $a=\min\{b_w\, |\, t\leq w\leq x-1\}$
\item[$(b)$] $x=B_{x-1}=\max\{B_w\, |\, t\leq w\leq x-1\}$
\item[$(c)$] for each $k$, $2\leq k\leq K$, the interval $I_k$ of $P_k$ delimited by the leftmost and rightmost values among $t, t+1, \ldots, x$ on $P_k$
satisfies  $\mini(I_k)\geq a$ and $\maxi(I_k)~=~x$.
\end{enumerate}
\label{fait:intLR}

\end{fait}

\noindent{\bf Proof of Claim \ref{fait:intLR}.} We use induction on $t$, with decreasing values. 
Let $t=n-1$ and notice that the $\LR$-stack is empty up to step 7 in {\bf for}$_{n-1}$.

\medskip
\noindent {\sf Proof of "$\Rightarrow$:"} The only operation that can affect the empty stack is $\PushLR(b_{n-1},n)$,
 performed only when $B_{n-1}=n$ (which is necessarily the case). Then $\Ra^{n-1}(b_{n-1})=\{n\}$. Thus $a=b_{n-1}, x=n$ and 
affirmations $(a), (b)$ are verified.  Now, the interval $I_k$ delimited by $n-1$ and $n$ satisfies the
required conditions since $\mini(I_k)=m_{n-1}\geq b_{n-1}=a$ and $\maxi(I_k)=n=x$.

\noindent{\sf Proof of "$\Leftarrow$:"} Let $a, x$ and $I_k$ be defined as in $(a)$-$(c)$. Because of 
$t=n-1$, together with $x\leq n$ (since $x$ is an element of $\n$) and with $t\leq w\leq x-1$ in $(b)$,  
we deduce that $x=n$ and $w=n-1$.
Thus $a=b_{n-1}$. 
By affirmation $(b)$, $B_{n-1}=n$. In conclusion, we have $b_{n-1}=a$ and $B_{n-1}=n$ 
thus the conditions in step 7 are fulfilled. Then $a$ is pushed on $L$, $n$ is pushed on $R$ 
and we are done. 
\bigskip

\noindent Assume now the claim is true for $t+1$, where $t+1\leq n-1$, and let us prove it for $t$. For that, 
we denote $a' :=\min\{b_w\, |\, t+1\leq w\leq x-1\}$.

\medskip

\noindent{\sf Proof of "$\Rightarrow$:"} With $x\in\Ra^{t}(a)$, three cases are possible:

\begin{enumerate}
\item[(i)] $x\in \Ra^{t+1}(a)$. In this case, during \fort\, $a$ is not discarded by $\PopL(\bt)$,
thus (1) $a\leq \bt$. Moreover, by the inductive hypothesis (affirmation $(a)$
for $t+1$), $x\in \Ra^{t+1}(a)$ implies that (2)~$a=\min\{b_w\, |\, t+1\leq w\leq x-1\}=a'$. Then 
by (1) and (2):

\begin{equation*}
\min\{b_w\, |\, t\leq w\leq x-1\}=\min\{\bt,a\}=a
\end{equation*}

\noindent and affirmation $(a)$ holds for $t$.  Furthermore, $x$ is 
not discarded by $\PopR(\Bt)$, thus (3) $x\geq \Bt$. Affirmation $(b)$ follows using the inductive hypothesis. Now, let  
$I'_k$ be the interval given by affirmation $(c)$ for $t+1$ and $P_k$.  Let 
$I''_k$ be the interval of $P_k$  delimited by $t$ and $t+1$. Defining $I_k=I'_k\cup I''_k$, we have that 
$I_k$ is an interval of $P_k$, since both $I'_k$ and $I''_k$ contain $t+1$. Also, $I_k$ is delimited 
as required, as $I'_k$ and $I''_k$ do. Moreover, using the hypothesis $\bt\leq \mint\leq \mint^k$ and $\Bt\geq \maxt\geq \maxt^k$, properties (1)-(3) above and the inductive hypothesis on $I'_k$ we obtain:

\begin{equation*}
\mini(I_k)=\min\{\mini(I'_k),\mini(I''_k)\}\geq \min\{a',\mint^k\}\geq \min\{a,\mint\}\geq \min\{a, \bt\}=a
\end{equation*}
\begin{equation*}
\maxi(I_k)=\max\{\maxi(I'_k),\maxi(I''_k)\}=\max\{x,\maxt^k\}=x.
\end{equation*}

\item[{(ii)}] there exists $a''> a$ such that $x\in \Ra^{t+1}(a'')$.
Then $a''$ is necessarily discarded by $\PopL(\bt)$, thus $a''>\bt$ and $\Ra^{t+1}(a'')$ goes entirely 
into $\Ra^t(\bt)$. Then, we deduce $x\in \Ra^t(\bt)$
and therefore $x\in \Ra^{t}(a)\cap \Ra^{t}(\bt)$, which is impossible unless (4) $a=\bt$. 
Now, by the inductive hypothesis, $x\in\Ra^{t+1}(a'')$ implies by affirmation $(a)$ that 
$a''=\min\{b_w\, |\, t+1\leq w\leq x-1\}=a'$. Thus, recalling that $a< a''$ by
the hypothesis of case (ii), we have (5) $a'=a''> a$. With (4) and (5) we deduce:

\begin{equation*}
\min\{b_w\, |\, t\leq w\leq x-1\}=\min\{\bt, a''\}=\min\{\bt,a'\}= \min\{a,a'\}=a
\end{equation*}

\noindent and affirmation $(a)$ holds for $t$. Furthermore, $x$ is 
not discarded by $\PopR(\Bt)$, thus (6) $x\geq \Bt$. Affirmation $(b)$ follows using the inductive hypothesis. 
As before, let  
$I'_k$ be the interval given by affirmation $(c)$ for $t+1$ and $P_k$. Let 
$I''_k$ be the interval of $P_k$  delimited by $t$ and $t+1$. Defining $I_k=I'_k\cup I''_k$, we have again that 
$I_k$ is an interval of $P_k$ delimited 
as required. Moreover, using the hypothesis $\bt\leq \mint\leq \mint^k$ and $\Bt\geq \maxt\geq \maxt^k$, properties (4)-(6) above and the inductive hypothesis on $I'_k$:

\begin{equation*}
\mini(I_k)=\min\{\mini(I'_k),\mini(I''_k)\}\geq \min\{a'',\mint^k\}\geq \min\{a,\mint\}\geq \min\{a, \bt\}=a
\end{equation*}
\begin{equation*}
\maxi(I_k)=\max\{\maxi(I'_k),\maxi(I''_k)\}=\max\{x,\maxt^k\}=x.
\end{equation*}

\item[{(iii)}] there is no $a''$ such that $x\in \Ra^{t+1}(a'')$.
In this case, $x$ is added to $R$ during \fort,
in step 8. Then $x=t+1=\Bt$, $a=\bt$ and affirmations $(a)$ and $(b)$ clearly hold.  
Recall that, by definition, $t+1\leq  \maxt^k\leq \maxt\leq \Bt$ thus $\maxt^k=t+1=x$.
Now, the interval  $I_k$ delimited by $t$ and $t+1$ on $P_k$ satisfies:

\begin{equation*}
\mini(I_k)\geq \mint\geq \bt=a
\end{equation*}
\begin{equation*}
\maxi(I_k)=\maxt^k=t+1=x
\end{equation*}

\noindent The "$\Rightarrow$" part of the claim is proved.
\end{enumerate}

\noindent{\sf Proof of "$\Leftarrow$:"} Let $a,x$ and $I_k$, $2\leq k\leq K$, be defined as in affirmations 
$(a)$-$(c)$ in the claim. Consider the two following cases:

\begin{enumerate}

\item[(i)] $t<x-1$. Notice that $a',x$ and the intervals $I'_k$, $2\leq k\leq K$, delimited on $P_k$ by the leftmost  and rightmost  values between $t+1,\ldots, x$ satisfy affirmations $(a)$-$(c)$ for 
$t+1$, thus by the inductive hypothesis $x\in\Ra^{t+1}(a')$. Now, we show that in both cases
occurring during \fort, we have (7) $\Ra^{t+1}(a') \subseteq \Ra^t(a)$. 
Indeed, when $a'$ is not discarded by $\PopL(\bt)$, we necessarily have $a'\leq \bt$ and 
thus we also have by affirmation $(a)$  that  $a=\min\{b_t, a'\}=a'$. Property (7) follows. 
When $a'$ is discarded, we necessarily have $\bt<a'$ and thus, by affirmation $(a)$, 
we deduce $a=\min\{b_t, a'\}=\bt$. The execution of  $\PopL(\bt)$ implies, in this case, 
that $\Ra^{t+1}(a')$ becomes a part of $\Ra^{t}(a)$, and (7) is proved.
Now, with (7) and given that, in step 6, $x=B_{x-1}\geq \Bt$ implies that $x$ is not 
discarded by  $\PopR(\Bt)$, we obtain that $x\in \Ra^t(a)$.

\item[(ii)] $t=x-1$. Then by hypothesis $a=b_{x-1}, x=B_{x-1}, \mini(I_k)\geq b_{x-1}$ and $\maxi(I_k)=x$.
In step 5 of {\bf for}$_{x-1}$, the instruction $\PopL(b_{x-1})$ discards all
$a''>b_{x-1}$ (if any). Thus, at the end of step 5, $\tp(L)\leq b_{x-1}$. The instruction $\PopR(x)$
insures that, at the end of step 6, $\tp(R)> x$. (Notice that $x=t+1$ and all the elements pushed before
on $R$ by $\PushLR$ are of the form $t'+1$ with $t'>t$). Then, the instruction $\PushLR(b_{x-1}, x)$
pushes $b_{x-1}$ on $L$ if necessary, pushes $x$ on $R$ (thus $\tp(L)=b_{x-1}$ and $\tp(R)=x$)
and adds $x$ to $\Ra^{t}(b_{x-1})$.
\end{enumerate}

\noindent Claim \ref{fait:intLR} is now proved. $\Box$
\bigskip

\noindent{\bf Proof of Theorem \ref{thm:B}.} By definition, $ (t..x)\in \st$ iff $x\in\Ra^t(t)$.
According to Claim \ref{fait:intLR}, this holds iff affirmations $(a)$-$(c)$ hold with $a=t$,
that is, the following affirmations hold simultaneously:

\begin{enumerate}
\item[$(a')$] $t=\min\{b_w\, |\, t\leq w\leq x-1\}$
\item[$(b')$] $x=B_{x-1}=\max\{B_w\, |\, t\leq w\leq x-1\}$
\item[$(c')$] for each $k$, $2\leq k\leq K$, the interval $I_k$ of $P$ delimited by the leftmost and rightmost values between 
$t, t+1, \ldots, x$ satisfies $\mini(I_k)\geq t$ and $\maxi(I_k)~=~x$.
\end{enumerate}

Now we show that $(a')-(c')$ hold iff $(t..x)$ is a common interval of $\PK$ (ignoring the signs) satisfying the conditions
in Theorem  \ref{thm:B}.

''$\Rightarrow$:'' We show that $I_k$ has the same elements as $(t..x)$ and that $\bt=t$ ($B_{x-1}=x$ is
directly given by affirmation $(b')$). By affirmation $(c')$, every element in $(t..x)$ is an
element of $I_k$. Conversely, assume by contradiction that $u$ is any element of $I_k$ 
distinct from $t, t+1, \ldots, x$. Then $u$ is not a delimiter of $I_k$. Consequently,
let $t'$, $t\leq t'\leq x-1$, such that  $u$ is between $t'$ and $t'+1$ on $P_k$. Then $b_{t'}\leq m_{t'}\leq 
u\leq M_{t'}\leq B_{t'}$.  
By affirmations $(a')$ and $(b')$, $t\leq b_{t'}$ and $B_{t'}\leq B_{x-1}=x$. Thus $t\leq u\leq x$,
a contradiction.
To show that $\bt=t$, notice that $\bt\leq t$, by the definition of $\bt$, and $t\leq \bt$ 
by affirmation $(a')$. 

''$\Leftarrow$:'' If $(t..x)$ is a common interval with $t=\bt=\min\{b_w\, |\, t\leq w\leq x-1\}$ and 
$x=B_{x-1}=\max\{B_w\, |\, t\leq w\leq x-1\}$, then obviously $(a')-(c')$ hold.
$\Box$

\section{Complexity issues}\label{sect:complex}

We separately discuss the implementation of an $\LR$-stack, and the running time of the $\LR$-Search algorithm.

\subsection{The $\LR$-stack}
The efficient implementation of an LR-stack depends on the need (or
not) to implement $\FindL$.  If $\FindL$ is not needed, then $L$ and $R$ may be  implemented
as lists. Consequently,  $\PopL$ and $\PopR$ are easily implemented in linear time with respect to the number
of elements removed respectively from $L$ and $R$, whereas $\PushLR$ takes constant time.
Also, $\tp(R), \tp(L)$ and  $\next()$  need $O(1)$ time.

When $\FindL$ is needed, then we are in the context of a Union-Find-Delete structure,
where the operations are performed on the  sets $\Ra(a)$, as follows: unions are 
performed by $\PopL$ and $\PushLR$, whereas deletions are performed by $\PopR$. These algorithms are already very efficient in the most general case \cite{Alstrup}, but unfortunately not linear. 
Yet, particular linear cases may be found and show useful (see Algorithm \ref{algo:max}). 

\subsection{The $\LR$-algorithm}

We prove the following result. 

\bthm Assume that computing $\bt$ and $\Bt$ takes negligible time and space with respect to the computation
of $\mint^k, \mint, \maxt^k$ and $\maxt$ in steps 1-2.  Then the $\LR$-Search algorithm has running  
time $O(Kn+F)$ and uses $O(n+f)$ additional space, where $F$ and $f$ respectively denote 
the running time and additional space needed by the \out\, procedure, over all values of $t$.
\label{thm:complex}
\ethm

\noindent{\bf Proof of Theorem \ref{thm:complex}.} Note that the $\FindL$ operation on the LR-stack is not needed in the
algorithm. Therefore, the LR-stack may be easily implemented so as to ensure linear running
times for $\PopL(\bt)$ and $\PopR(\Bt)$ with respect to the number $dl_t$ and $dr_t$ of discarded elements,
and constant running time for $\PushLR$.  Then the {\bf for} loop in steps 4-11 takes
running time $O(\Sigma_{1\leq t<n}(dl_t+dr_t+F_t))$, where $F_t$ is the running time of the \out\, procedure for $t$, that is $O(\Sigma_{1\leq t<n}(dl_t+dr_t)+F)$.
Furthermore, each variable $\bt$ and $\Bt$ is pushed on the $\LR$-stack at most once
(step 8), and thus it is discarded from the $\LR$-stack at most once. Consequently,
$\Sigma_{1\leq t<n}(dl_t+dr_t)$ is in $O(n)$ and the {\bf for} loop takes $O(n+F)$ total time.
Concerning the memory requirements, it is clear that $L$ and $R$ are filled in with
elements $\bt,\Bt$ whose cardinality is in $O(n)$.

Given the hypothesis that computing the pairs $[\bt,\Bt]$ for $t\in [n-1]$ takes negligible
time and space, it remains to show that the other computations in steps 1 and 2 take  $O(Kn)$ time and  $O(n)$  
additional space. To this end, we compute $\mint^k$ and $\maxt^k$, $t\in[n-1]$, for 
each permutation $P_k$ in $O(n)$ time and $O(n)$ additional space, as described below.
The values computed for each permutation are progressively 
included in the  computation of  $\mint, \maxt$, so as to use a global $O(n)$ space.
 
In \cite{berkman1993recursive, bender2000lca}, authors solve a problem called range minimum query 
(abbreviation: RMQ problem). More precisely, they show that, given any array $A$ of $n$ numbers, 
it is possible to preprocess it in $O(n)$ time so as to answer in $O(1)$ any query asking for (the position of) the
minimum value between two given positions $q_1$ and $q_2$ in $A$. This 
result, closely related to computing the least common ancestor of two given nodes 
in a rooted tree, allows us to compute $\mint^k, \maxt^k$ for all $t$ in 
linear time, for each permutation $P_k$. Then we are already done.

However, we propose here another algorithm, answering a set $Q$ of queries in $O(n+|Q|)$ time.
This algorithm is obviously less powerful than the  preceding ones, but has at least
two advantages. First, it is conceptually and algorithmically simpler, allowing the
reader to immediately simulate executions. Second, it gives another application of 
$\LR$-stacks, needing this time the implementation of the $\FindL$ operation.

Algorithm ComputeInf is given in  Algorithm \ref{algo:max}.
The input is an arbitrary permutation $P=(p_1, p_2,$ $\ldots, p_n)$, signed or not, to
which the algorithm adds an element $p_0=n+1$.
It is quite easy to notice that the algorithm works similarly for an array,
i.e. when elements are not unique.

For a pair $(q_1,q_2)$ with $0\leq q_1<q_2\leq n$, denote 

$$\up(q_1,q_2)=\min\{p_h\,|\, q_1\leq h\leq q_2\}.$$

%
\smallskip

Algorithm ComputeInf computes $\up(q_1,q_2)$ for all pairs in some given set $Q$, in
$O(n+|Q|)$ time. Intuitively, the variable $h$ in step 3 considers all possible values $q_2$, stack $R$
contains all possible values of $q_1$, whereas stack $L$ contains in $\FindL(q_1)$ the best
current candidate for $\up(q_1,q_2)$. 

\begin{fait}
Let $h\in\n\cup\{0\}$. After the execution of {\bf for}$_h$ in step 3, the $\LR$-stack satisfies
the property $i\in\Ra(a)$ iff $\up(i,h)=a$.
\label{fait:up}
\end{fait}

\noindent{\bf Proof of Claim \ref{fait:up}.} Notice that only allowed operations are performed
on the $\LR$-stack, except that $\PushLR(p_{h},h)$ is not preceded by $\PopR(h)$.
This operation would have no effect, since at the beginning of the $h$-th step all the elements 
in $R$ are already less than $h$. 

Remark that in a $\LmRm$-stack, $a'$ is $L$-blocking for $a$ iff $a'>a$, and $b'$ is $R$-blocking
for $b$ iff $b'>b$. 
We use induction. For $h=0$, at the beginning of the execution the $\LR$-stack is empty, and
only $\PushLR(n+1,0)$ is executed, insuring the claim is true. Assuming
the claim true for $h$, let us show it is true for $h+1$. 
\medskip

\noindent{\sf Proof of "$\Rightarrow$".} Let $i\in \Ra(a)$ at the end of the execution  {\bf for}$_{h+1}$. Two cases may appear:

\begin{enumerate}

\item[(i)] When $a<p_{h+1}$, we deduce that $\PopL(p_{h+1})$ did not discard $a$, thus $i\in \Ra(a)$ 
at the end of the preceding execution of the {\bf for} loop. By the inductive hypothesis, 
we deduce that $\up(i,h)=a$ and thus, with $a<p_{h+1}$, we have that $\up(i,h+1)=a$.

\item[(ii)] When $a=p_{h+1}$, either $i$ has been added by $\PushLR(p_{h+1},h+1)$, or some
$a'>p_{h+1}$ existed on $L$ before $\PopL(p_h)$ such that $i\in\Ra(a')$ at the end of the execution
of the {\bf for} loop for $h$. In the first case, $i=h+1$ and the conclusion obviously holds.
In the second case, the inductive hypothesis implies that $\up(i,h)=a'$. Consequently, 
we have that $\up(i,h+1)=p_{h+1}=a$, since $a'>p_{h+1}$.

\end{enumerate}

\begin{algorithm}[t,boxed]
\caption{The ComputeInf algorithm}
\begin{algorithmic}[1]
\REQUIRE Permutation $P$ over $\n$, $Q\subseteq\{(q_1,q_2)\, |\, 1\leq q_1<q_2\leq n\}$
\ENSURE  Values $\up(q_1,q_2)$ for all $(q_1,q_2)\in Q$
\STATE Initialize an empty $\LmRm$-stack
\STATE $p_0\leftarrow n+1$
\FOR{$h$ from $0$ to $n$}
\STATE $\PopL(p_{h})$ {\sl \hfill //update $\up(i,h)$ for all $i< h$ s.t. $\up(i,h-1)>\up(i,h)$}
\STATE $\PushLR(p_{h},h)$ {\sl \hfill // $\up(h, h)=p_h$}   
\FOR{all pairs $(q_1,q_2)\in Q$ such that $h=q_2$}
\STATE $\query(q_1,q_2)\leftarrow \FindL(q_1)$ 
\ENDFOR
\ENDFOR
\end{algorithmic}
\label{algo:max}
\end{algorithm} 

\noindent {\sf Proof of "$\Leftarrow$".} By the hypothesis, we assume $\up(i,h+1)=a$ at the end of the
execution of the {\bf for} loop for $h+1$.

\begin{enumerate}
\item[(i)] When $a<p_{h+1}$, we deduce $\up(i,h)=a$ and by the inductive hypothesis we obtain
$i\in\Ra(a)$ at the end of the execution of the {\bf for} loop for $h$. Since $a<p_{h+1}$,
the instruction $\PopL(p_{h+1})$ does not discard $a$ and the conclusion follows.
\item[(ii)] When $a=p_{h+1}$, we deduce $\up(i,h)>p_{h+1}$, since $p_{h+1}$ occurs only once on $P$. By the inductive hypothesis, that means
$i\in\Ra(a')$ at the end of the execution of the {\bf for} loop for $h$, with $a'=\up(i,h)>p_{h+1}$. Consequently, 
$\PopL(p_{h+1})$ discards $a'$ and moves $i$ into $\Ra(p_{h+1})$ (which is $\Ra(a)$).

\end{enumerate} 

The claim is proved.$\Box$

\begin{fait} For each $(q_1,q_2)\in Q$, the value $\query(q_1,q_2)$ returned by Algorithm ComputeInf 
is equal to $\up(q_1,q_2)$. Moreover, the algorithm may be implemented to have $O(n+|Q|)$ running time.
\label{fait:upcorrect}
\end{fait}

\noindent{\bf Proof of Claim \ref{fait:upcorrect}.} The value $\query(q_1,q_2)$ is computed 
in step 7 of the {\bf for} loop, when $h=q_2$. By Claim \ref{fait:up}, $\FindL(q_1)=\up(q_1, q_2)$
and we are done.

Concerning the running time, it is easy to see that step 6 (we temporarily leave apart step 7) 
may be performed in $O(n+|Q|)$ over all the values of $h$, by sorting the pairs $(q_1,q_2)$ in $Q$ 
according to the lexicographic order of  the pairs $(q_2,q_1)$ (for instance using radix sort). 
Concerning step 7, the main difficulty 
comes from the need to implement the 
operation $\FindL$. However, we benefit here from a very particular case of the 
Union-Find-Delete context, where no deletion is performed (i.e. no $\PopR$ is performed) and the union
operations (due to $\PopL$ and $\PushLR$) always join
sets of consecutive elements to obtain another set of consecutive elements. In \cite{itai2006linear},
an implementation of the Union-Find operations for this particular case is proposed, which
realizes each union between two sets in $O(1)$, and $m$ operations $Find$ in $O(n+m)$.
Moreover, the sets (equivalently: the elements of $L$) as well as their elements (equivalently:
the elements in $R$) may be easily chained to simulate the $L$ and $R$ stacks. 
Then each $\PopL$ operation takes linear time with respect to the number of discarded 
elements and is in $O(n)$ time over all $h$, since the elements on $L$ are elements of $P$, 
which are pushed exactly once on $L$.  Furthermore, $\PushLR$ takes constant time and
the $|Q|$ calls of $\FindL$ take $O(n+|Q|)$ time. The 
indicated running time for our algorithm follows. $\Box$
\medskip

\noindent{\bf Proof of Theorem \ref{thm:complex} (continued).} Algorithm ComputeSup, which
computes the maximum values between given pairs of positions in $P$, is clearly similar to ComputeInf. Then, 
$\mint^k$ and $\maxt^k$ may be obtained by appropriately defining the query set $Q$, and the running time of $\LR$-Search follows.$\Box$

\section{Finding common, nested and conserved intervals}\label{sect:comm}

In this section and the next one, we prove the following general theorem. The 
definitions of the subclasses not yet defined are provided later, just before they are used.

\begin{thm} {\bf (Unifying Theorem)}
Algorithm $\LR$-Search with settings $\BL, \BR$ and \out$\,$ given in Table \ref{tab:unifying} 
solves $\C$-\ISP\, for $K$ permutations in $O(Kn+N)$ time and $O(n)$ additional space, where 
\medskip

$\C\in \{$Common, Nested, Conserved,
Irreducible Common, Same-Sign Common, 

\hspace*{7cm} Maximal Nested, Irreducible Conserved$\}$
\medskip

\noindent and $N$ is the number of $\C$-intervals of $\PK$.
\label{thm:unify}
\end{thm}

{\small 
\begin{table}
\begin{center}
\begin{tabular}{|l||c|c|}
 \hline

{\bf Class $\C$}& $\BL, \BR$ {\bf for} $i$, $1\leq i\leq n-1$ & {\bf \out$\,$}\\ \hline\hline
Common&$\BL(i) :=m_i$, $\BR(i) :=M_i$ & Alg. \ref{algo:filtercommon}\\\hline
Nested& $\BL(i) : =m_i$, $\BR(i) :=M_i$& Alg. \ref{algo:filternested}\\ \hline
Conserved&$\BL(i) :=\min\{u^k_i\, |\, 2\leq k\leq K\}$ & Alg. \ref{algo:filterconserved} \\ 
&$\BR(i) :=\max\{v^k_i\, |\, 2\leq k\leq K\}$&\\
&&\\
& with $u^k_i := \left\{\begin{array}{ll} m^k_i & \mbox{if}\, m^k_i=i \\ 
                                        m^k_i-1 & \mbox{otherwise} \end{array}\right. $&\\
&\hspace{1.5cm}$v^k_i :=\left\{\begin{array}{ll} M^k_i &\mbox{if}\,  M^k_i=i+1\\
																				M^k_i+1 & \mbox{otherwise}\end{array}\right.$ &\\ \hline
Irreducible Common&  same as Common& Alg. \ref{algo:filterirr}\\  \hline
Same-Sign Common& same as Common& in text \\ \hline
Maximal Nested& same as Nested&Alg. \ref{algo:filternested}+\ref {algo:filternested2}\\ \hline
Irreducible Conserved& same as Conserved&in text \\  \hline

\end{tabular}
\end{center}
\caption{Classes $\C$ for which Algorithm $\LR$-Search solves $\C$-\ISP.}
\label{tab:unifying}
\end{table}}

For the ease of presentation, we assume that  $\Rb(a)$ is a  pointer to
 the last element of $\Ra(a)$, for all $a\in \SL$ with
$\Ra(a)\neq\emptyset$. It is easy to check that such a pointer is easily updated during
the operations on the $\LR$-stack.
 
\br Note that, according to the preceding definitions, the set $\Ra(t)$ is, at the
end of step 9 of \fort in the $\LR$-Search algorithm, the set denoted $\Ra^t(t)$.
\label{rmk:cond}
\er

The three subsections of this section respectively concern common, nested and conserved intervals.

\subsection{Common intervals}\label{subsect:common}

In this case, $\PK$ is a set of permutations (all elements have a + sign). The $\M$-profile of $\PK$ uses in this
case the basic settings for $\BL$ and $\BR$:

\begin{enumerate}
\item[$\bullet$] $\BL(i) :=m_i$, for all $i$, $1\leq i\leq n-1$
\item[$\bullet$] $\BR(i) :=M_i$, for all $i$, $1\leq i\leq n-1$
\item[$\bullet$] \out$\,\,$is given in Algorithm \ref{algo:filtercommon}. 
\end{enumerate}

\begin{algorithm}[t,boxed]
\caption{The \out$\,\,$algorithm for common intervals} 
\begin{algorithmic}[1]
\REQUIRE Pointers $\Rt(t),\Rb(t)$ to the first and last element of $\Ra(t)$ (possibly equal to $nil$)
\ENSURE  All common intervals $(t..x)$ of $\PK$, with fixed $t$.

\IF{$\Rt(t)\neq nil$}
\STATE $x^{\top}\leftarrow $ the target of $\Rt(t)$; $x^{\bot}\leftarrow $ the target of $\Rb(t)$
\STATE $x\leftarrow x^{\top}$
\WHILE{$x\leq x^{\bot}$}
\STATE Output the interval $(t..x)$
\STATE $x\leftarrow next(x)$ \hfill {\sl //or $n+1$ if $\next(x)$ does not exist}
\ENDWHILE
\ENDIF
\end{algorithmic}
\label{algo:filtercommon}
\end{algorithm}

\bex
Recall that, in Figure \ref{fig:LRSearch}, $\PK=\{P_1,P_2\}$ with $P_1=\Id_7$ and 
$P_2=(7\, 2\, 1\,3\, 6\, 4\, 5)$. The \out\, procedure in Algorithm \ref{algo:filtercommon}
successively outputs the intervals $(4..5),(4..6)$ (when $t=4$), $(3..6)$ (when $t=3$) and $(1..2), (1..3)$, $(1..6)$, $(1..7)$ (when $t=1$).
\eex

\begin{thm}
Algorithm $\LR$-Search with settings $\BL, \BR$ and \out$\,$ above solves Common-\ISP\, for $K$ permutations in $O(Kn+N)$ time and $O(n)$ additional space,
where $N$ is the number of common intervals of $\PK$.
\label{thm:common}
\end{thm}

\noindent{\bf Proof of Theorem \ref{thm:common}.} By Remark \ref{rmk:cond}, Algorithm \ref{algo:filtercommon} considers
each $x$ in $\Ra^t(t)$. Thus, over all values of $t$,  \out$\,\,$outputs $\cup_{1\leq t\leq n-1}\st$. 
By Theorem \ref{thm:B}, this is exactly the set of common intervals $(t..x)$ of $\PK$ with  
$t=\bt=\min\{b_w\,|\, t\leq w\leq x-1\}$ and 
$x=B_{x-1}=\max\{B_w\,|\, t\leq w\leq x-1\}$. Thus all the intervals output by the algorithm are common intervals of $\PK$.
Conversely, let $(t..x)$ be a common interval of $\PK$. Then, for each $k$, $2\leq k\leq K$,
and each $w$, $t\leq w\leq x-1$, we must have $t\leq m_{w}^k\leq w<w+1\leq M_{w}^k\leq x$.
Consequently, $t\leq m_w\leq w< w+1\leq M_w\leq x$ for all $w$, $t\leq w\leq x-1$.
By definition, $b_i=\BL(i)=m_i$ and $B_i=\BR(i)=M_i$ for all $i$, and thus we have 
$t\leq b_w$ and $B_w\leq x$ for all $w$. We deduce that

$$\bt\leq t\leq \min\{b_w\,|\, t\leq w\leq x-1\}$$

$$B_{x-1}\geq x \geq \max\{B_w\,|\, t\leq w\leq x-1\}$$

\noindent and thus $t=\bt=\min\{b_w\,|\, t\leq w\leq x-1\}$ and 
$x=B_{x-1}=\max\{B_w\,|\, t\leq w\leq x-1\}$. By Theorem \ref{thm:B} we obtain that $(t..x)$
is output by $\LR$-Search and the correctness of the algorithm is proved.

The {\bf while} loop in steps 4-7 of \out\, takes $O(|\Ra^t(t)|)$ time, since $\Rt(t)=\tp(R)$. \out$\,\,$has therefore a linear complexity with respect to the size of the output.
By Theorem \ref{thm:complex} and given that $\BL(i)$ and $\BR(i)$ are computed in constant time for each $i$
when $m_i$ and $M_i$ are known, the $\LR$-Search algorithm with the abovementioned settings runs in
$O(n+N)$ time. The additional space used by \out\, is in $O(1)$.$\Box$

\subsection{Nested Intervals}\label{subsect:nested}

In this case too, $\PK$ is a set of permutations (all elements have + sign) and the $\M$-profile
uses the basic settings for $\BL$ and $\BR$:

\begin{enumerate}
\item[$\bullet$] $\BL(i) : =m_i$, for all $i$, $1\leq i\leq n-1$
\item[$\bullet$] $\BR(i) : =M_i$, for all $i$, $1\leq i\leq n-1$
\item[$\bullet$] \out$\,\,$is given in Algorithm \ref{algo:filternested}. 
\end{enumerate}

\bex
Again, in Figure \ref{fig:LRSearch} the $\LR$-Search algorithm is applied to 
$\PK=\{P_1,P_2\}$, where $P_1=\Id_7$ and $P_2=(7\, 2\, 1\, 3\, 6\, 4\, 5)$.
When $t=4$, the \out\, procedure in Algorithm \ref{algo:filternested} outputs
intervals $(4..5), (4..6)$ and sets $W[3]\leftarrow 6$. When $t=3$, $\tp(L)= 3$
and $W[3]\neq 0$, so that $(3..6)$ is output and $W[2]$ receives the value 6.
Now, when $t=2$ we have $\tp(L)\neq 2$ and the {\bf if} instruction in step 4 stops 
the execution of \out. Then $W[1]$ remains 0. Thus, when $t=1$, in step 9 of \out\,
we have $x^{\top}=t+1$ and the algorithm outputs  $(1..2), (1..3)$ but not 
$(1..6)$, nor $(1..7)$.
\eex

In this subsection, we say that $(x,x+1)$ is a {\em gap} if exactly one
of $x,x+1$ is on $R$. 

\begin{algorithm}[t,boxed]
\caption{The \out$\,\,$algorithm for nested intervals}
\begin{algorithmic}[1]
\REQUIRE Pointers $\Rt(t),\Rb(t)$ to the first and last element of $\Ra(t)$ (possibly equal to $nil$)
\ENSURE  All nested intervals $(t..x)$ of $\PK$ with fixed $t$.

\IF{$t=n-1$}
\STATE Let $W$ be a $n$-size vector filled in with 0
\ENDIF 
\IF{$\Rt(t)\neq nil$}
\STATE $x^{\top}\leftarrow $ the target of $\Rt(t)$; $x^{\bot}\leftarrow $ the target of $\Rb(t)$
\IF{$x^{\top}>W[t]$} 
\STATE  $W[t]\leftarrow 0$ {\sl \hfill // element $W[t]$ has been discarded by $\PopR(\Bt)$}
\ENDIF
\IF{$x^{\top}=t+1$ or $W[t]\neq 0$} 
\STATE $x\leftarrow x^{\top}$
\WHILE{$x\leq x^{\bot}$ and ($x=t+1$ or $x\leq W[t]$ or $(x-1,x)$ is not a gap)}
\STATE Output the interval $(t..x)$
\STATE $W[t-1]\leftarrow x$ {\sl \hfill // $t$ passes down to $t-1$ its largest $x$ such that $(t..x)$ is output}
\STATE $x\leftarrow\next(x)$ {\sl \hfill //or $n+1$ if $\next(x)$ does not exist}
\ENDWHILE
\ENDIF
\ENDIF
\end{algorithmic}
\label{algo:filternested}
\end{algorithm}

\begin{thm}\label{thm:nested}
Algorithm $\LR$-Search with settings $\BL, \BR$ and \out$\,$ above solves Nested-\ISP\, for $K$ 
permutations in $O(Kn+N)$ time and $O(n)$ additional space, where $N$  is the number of nested intervals of $\PK$.

\end{thm}

By Remark \ref{rmk:cond}, $\Ra(t)$ in Algorithm \ref{algo:filtercommon} is the
same as $\Ra^t(t)$. Notice that the value of $W[t]$ computed by Filter during its execution for $t+1$ indicates the largest 
element $x$ in $\Ra^{t+1}(t+1)$ such that $(t+1..x)$ is output by \out$\,\,$(step 13 in \out$\,\,$for $t+1$). 
However, $W[t]$ may be set to $0$ in step 7 of the \out$\,\,$procedure (when executed for $t$),
if we know that its initial value was discarded by $\PopR(\Bt)$ in the $\LR$-Search algorithm.

From now on, we focus on the execution of the {\bf for} loop in $\LR$-Search for $t$
(including \out).
Let $y_t$ be defined as follows. We use the notations in step 5 of the \out$\,\,$procedure,
and assume thus that $\Rt(t)\neq nil$. 
If $x^{\top}=t+1$ or $W[t]\neq 0$, then $y_t$ is the 
first element in  $\Ra^t(t)$ such that the following properties hold:
\smallskip

(1) $W[t]\leq y_t$, and 

(2)  either $(y_t,y_t+1)$ is a gap, or $y_t=x^{\bot}$. 
\smallskip

Otherwise, {\it i.e.} if  $x^{\top}\neq t+1$ and $W[t]= 0$, $y_t$ is  set to 0.  Let 

\begin{equation*}
V_t :=\{v\in \Ra^t(t)\,|\, x^{\top} \leq v \leq y_t\}.
\end{equation*}

Now, let us prove that:

\begin{fait} 
The intervals output by \out$\,$ are  the intervals $(t..x)$ with $x\in V_t$. 
\label{fait:Vt}
\end{fait}

\noindent{\bf Proof of Claim \ref{fait:Vt}.} Indeed, when 
$x^{\top}\neq t+1$ and $W[t]= 0$, the condition in step 9 is false and the
algorithm returns an empty set of intervals. This is correct, since $V_t=\emptyset$ in this
case. 

In the contrary case, the condition in step 9 is true. The {\bf while} loop starts 
with $x=x^{\top}$ and outputs all intervals  $(t..x)$ with $x$ in $\Ra^t(t)$ up
to a last one $(t..x_0)$. 
If $x_0=x^{\bot}$, then $x_0=x^{\bot}=y_t$, since $x^{\bot}$ satisfies properties (1) and (2) 
in the definition of $y_t$ whereas no other element preceding it in $\Ra^t(t)$ does. 
If $x_0\neq x^{\bot}$, then $x_0$ satisfies the second condition in step 11, whereas
the element $x'=\next(x)$ is in $\Ra^{t}(t)$ but does not satisfy the second condition in step 11.
Then $(x'-1,x')$ is a  gap and thus $(x_0,x_0+1)$ is a gap too. In order to 
deduce that $x_0=y_t$, we only have to show that condition (1) in the definition of
$y_t$ is satisfied by $x_0$. If, by contradiction, $W[t]>x_0$, then $W[t]$ is on $R$ below
$x_0$ and thus $x'\leq W[t]$. This is impossible, since $x'$ does not satisfy the
second condition in step 11.$\Box$
\medskip

The following claim establishes the correctness of our algorithm: 

\begin{fait}
We have $x\in V_t$ iff $(t..x)$ is a nested interval.
\label{fait:nest}
\end{fait}

\noindent{\bf Proof of Claim \ref{fait:nest}.}  This proof is by induction on $t$. For $t=n-1$,  $\Rt(n-1)\neq nil$ iff $\PushLR(\bt, t+1)$ is called in $\LR$-Search with
$\bt=n-1$ and $t+1=n$. Thus $\Ra^{n-1}(n-1)=\{n\}$.
Then, with $W[n-1]=0$, we have that $V_{n-1}=\{n\}$ and indeed $(n-1..n)$ is a nested interval. 

We now assume the claim is true for $t+1$ and show it for $t$. Then $\Rt(t)\neq nil$
in step 4 of \out\, (otherwise $V_t$ is not defined, and there is no nested interval with left
endpoint $t$) and thus  $\PopL(\bt)$
 in step 5 of $\LR$-Search has been executed with $\bt=t$. Consequently,
$V_{t+1}\subseteq \Ra(t)$ at the end of step 5 in \fort\, of $\LR$-Search. In step 6, 
all elements $x$ of $R$ with
$x<\Bt$ are discarded and, with them, the first elements of $V_{t+1}$ (since the
first element of $\Ra(t)$, as well as of $V_{t+1}$ if it is non-empty, is $\tp(R)$). 
Furthermore, the $\PushLR(\bt,t+1)$ operation, if executed,  pushes $t+1$ on the
top of $R$ and, in the same time, in $\Ra^t(t)$. 

When \out$\,\,$is executed for $t$, several situations may occur. Notations $x^{\top}$ and $x^{\bot}$
concern the execution of the {\bf  for} loop in $\LR$-Search, including \out, for $t$. Then $x^{\top}=\tp(R)$.

\begin{enumerate}

\item[(i)] $x^{\top}=t+1$ and $W[t]=0$ in step 9 of \out. Then either there is no $x$
such that $(t+1..x)$ is output, or such an $x$ exists but is removed from $R$ by $\PopR(\Bt)$.
According to the definition, $y_t$ is the first element of $\Ra^t(t)$ such that either
$(y_t,y_t+1)$ is a gap, or $y_t=x^{\bot}$. In both cases, since $x^{\top}=t+1$ we have
that $t+1$ is in $V_t$. Also, $(t..t+1)$ is
nested since it is a common interval (by Theorem \ref{thm:B}) and has cardinality 2. 
By the definition of nested intervals, all intervals $(t..x)$ with successive values of 
$x$, $x\geq t+2$, will be nested, thus $x\in V_t$ implies that $(t..x)$ is nested. Conversely,
assume by contradiction that some $x\geq t+2$ exists such that $(t..x)$ is nested but
$x\not\in V_t$, and take $x$ as small as possible with these properties. 
Since $(t..x)$ is nested, $(t..x)$ is a common interval thus, by
Theorems \ref{thm:B} and \ref{thm:common} and given the settings of $\BL$ and $\BR$, we have $x\in \Ra^t(t)$, thus $x\leq x^{\bot}$. As $(t..x)$ is nested, either $(t+1..x)$ or $(t..x-1)$ is nested. 
The former does not hold (because of $W[t]=0$), 
thus $(t..x-1)$ is nested. Since $x$ was the smallest with the indicated properties, $x-1\in V_t$.
But then, by the definition of $y_t$, we have $x\in V_t$ too, a contradiction.

\item[(ii)] $x^{\top}=t+1$ and $W[t]>0$ in step 9 of \out. Then

\begin{equation*}
V_t=\{t+1\}\cup V_{t+1}-\{v\in V_{t+1}\,|\, v <t+1\}\cup U
\label{eq:Vt}
\end{equation*}

\noindent where $U$ is the set of all consecutive elements $W[t]+1,W[t]+2, \ldots, z \subseteq 
\Ra^t(t)$  such that either $z=x^{\bot}$ or $(z,z+1)$ is a gap. Equivalently, $V_t$ gets all 
the elements in $V_{t+1}$ except those smaller than $t+1$ (because of $\PopR$), 
as well as $t+1$ (pushed by $\PushLR$) and all the consecutive elements that are possibly 
added at the end of $V_{t+1}$ during the $\PopL(\bt)$ operation, with $\bt=t$.  It is
easy to see that the last element of $U$ satisfies the conditions (1) and (2) in the
definition of $y_t$, and no
one before it in $\Ra(t)$ does. Thus $z=y_t$.

By contradiction, assume some $x\in V_t$ exists such that $(t..x)$ is not nested. Assume
$x$ is the smallest with these properties. Then $x>W[t]$, otherwise $x=t+1$ or $x\in V_{t+1}$
and then $(t+1..x)$ is nested by definition and the inductive hypothesis, insuring that $(t..x)$ is nested. Now, $x-1\in V_t$ (since there is
no gap in $V_t$ beyond $W[t]$) thus, by the minimality of $x$, $(t..x-1)$ is nested. But then
$(t..x)$ is nested, a contradiction. Conversely, let $(t..x)$ be a nested interval, and let us
show that $x\in V_t$. Once again, assume this is not true and let $x$ be the smallest counter-example.
Since $(t..x)$ is nested, $(t..x)$ is a common interval and, by Theorems \ref{thm:B} and \ref{thm:common}, 
$x\in\Ra^t(t)$ thus $x\geq t+1$. Now, we must have $x>W[t]$, otherwise $x=t+1$ or $x\in V_{t+1}$ 
thus $x\in V_t$, a contradiction. Finally, since $(t..x)$ is nested we have two cases.  
Either $(t+1,x)$ is 
nested, and then by the inductive hypothesis $x\in V_{t+1}$ thus $x\in V_t$, a contradiction.  Or 
$(t,x-1)$ is nested and thus $x-1\in V_t$ by the minimality of $t$, thus $x\in V_t$ by the
definition of $V_t$, another contradiction.

\item[(iii)] $x^{\top}\neq t+1$ and $W[t]>0$ in step 9 of \out. Then 

\begin{equation*}
V_t=V_{t+1}-\{v\in V_{t+1}\,|\, v< t+1\}\cup U
\end{equation*}

\noindent with $U$ defined as previously done, and the proof follows similarly.

\item[(iv)] $x^{\top}\neq t+1$ and $W[t]=0$ in step 9 of \out. Then $V_t=\emptyset$
and we must show there is no nested interval $(t..x)$. If, by contradiction,
such an $x$ exists, then assume $x$ is taken to be the smallest one. Then
$(t..x-1)$ is not nested, by the minimality of $x$. Thus $(t+1,x)$ is nested,
and thus $x\in V_{t+1}$. But then $x\in V_t$ unless $x$ is removed by $\PopR(\Bt)$,
with $\Bt=t+1$. However, this is impossible, since $x\geq t+1$.$\Box$ 
\end{enumerate}

\br
According to the preceding claim, $W[t]=y_{t+1}$. Moreover, assume that every element $r$ in
$R$ has an associated pointer $R^g(r)$ on the first element $w\in R$ larger than or equal to 
$r$ and such that either $(w,w+1)$ is a gap or $w$ is the bottom of $R$. Then $y_t$
may be computed in constant time in each of the cases (i)-(iv) of the proof, 
using $W[t]$ ({\it i.e.} $y_{t+1}$) and $R^g(r_0)$, where $r_0$ is the target of $\Rt(t)$ at the end of 
{\bf for}$_{t+1}$, if $\Rt(t)\neq nil$.
\label{rem:W}
\er

\noindent{\bf Proof of Theorem \ref{thm:nested}.} 
The algorithm correctness is proved by Claims \ref{fait:Vt} and \ref{fait:nest}.
The $O(Kn+N)$ running time of the algorithm is due to Theorem \ref{thm:complex}
and to the linearity of \out$\,\,$with respect to $|V_t|$, which is clear for $t<n-1$
since \out\,\, stops when the first element not in $V_t$ is found. The $O(n)$ complexity
when $t=n-1$ does not change the overall running time of the algorithm.$\Box$ 
\bigskip

\subsection{Conserved intervals}\label{subsect:conserved}

Here, each $P_k$ is a signed permutation with first element $1$ and last element $n$, both positive.
The definitions of $\BL$ and $\BR$ define a $\M$-profile of $\PK$ adapted to the
specific needs of conserved intervals.
For each $i$ with $1\leq i\leq n-1$, let $u^k_i := m^k_i$ if $m^k_i=i$, 
and $u^k_i :=m^k_i-1$ otherwise. Similarly, let $v^k_i :=M^k_i$ if $M^k_i=i+1$, 
and $v^k_i :=M^k_i+1$ otherwise.  Let:

\begin{enumerate}
\item[$\bullet$] $\BL(i) :=\min\{u^k_i\, |\, 2\leq k\leq K\}$,  for all $i$, $1\leq i\leq n-1$ 

\item[$\bullet$] $\BR(i) :=\max\{v^k_i\, |\, 2\leq k\leq K\}$, for all $i$, $1\leq i\leq n-1$
\item[$\bullet$] \out$\,\,$is given in Algorithm \ref{algo:filterconserved}, where: $\Rt(t)^*$ is a pointer to the first element $x$ in $\Ra(t)$ such that $t$ and $x$ have the same sign
in all permutations in $\PK$ (such elements $x$ are chained together inside $R$); and $\Position(t,t+1)$
returns true iff, for each $k$, either $t$ is positive in $P_k$ and $\Pm_k(t)<\Pm_k(t+1)$, or $t$ is negative in $P_k$ and $\Pm_k(t)>\Pm_k(t+1)$.
\end{enumerate}

\br Note that $(t..x)$ is a conserved interval of 
$\PK$ iff  it has the following properties: 

(1) it is a common interval of $\PK$

(2) it is delimited by $t$ and $x$ on $P_k$, for all $k\in[K]$

(3) $t$ and $x$ are both positive or both negative in each $P_k$, for all $k\in[K]$

(4) for each $k\in[K]$, either $t$ is positive in $P_k$ and $\Pm_k(t)<\Pm_k(x)$, or $t$ is
negative in $P_k$ and $\Pm_k(x)<\Pm_k(t)$.
\label{rmk:charcons}
\er

Conditions (1) and (2) are easily handled by defining the bounding functions $\BL$ and
$\BR$ as indicated. However, conditions (3) and (4) need a preprocessing of the permutations
in $\PK$, in order to: (Task 1) identify and chain together inside 
$R$ the elements $x$ in the same equivalence class with respect to the relation ``$x$ and $x'$ 
have the same sign in all permutations'', allowing us to deal with (3); and (Task 2) compute the boolean function $\Position()$ defined above, which will insure that (4) holds. These tasks
are done in $O(Kn)$ time and $O(n)$ additional space as follows:

\begin{algorithm}[t,boxed]
\caption{The \out$\,\,$algorithm for conserved intervals}
\begin{algorithmic}[1]
\REQUIRE Pointers $\Rt(t)^*$  to the first element in $\Ra(t)$
having the same sign as $t$ in all permutations $P_k$, and $R^{\bot}(t)$ to the last element in $\Ra(t)$ (they are possibly $nil$)
\ENSURE  All conserved intervals $(t..x)$ of $\PK$ with fixed $t$.

\IF{$\Rt(t)^*\neq nil$}
\STATE $x^{\top}\leftarrow $ the target of $\Rt(t)^*$; $x^{\bot}\leftarrow $ the target of $\Rb(t)$  
\STATE $x\leftarrow x^{\top}$
\IF{$Position(t,t+1)$}
\WHILE{$x\leq x^{\bot}$}
\STATE Output the interval $(t..x)$
\STATE $x\leftarrow$ the element immediately following $x$ in its chain \hfill{ \sl //or $n+1$ if it does not exist}
\ENDWHILE
\ENDIF
\ENDIF

\end{algorithmic}
\label{algo:filterconserved}
\end{algorithm}

{\bf Task 1.} Consider the matrix $M$ whose
row vectors are the vectors $\sign_{P_k}$ for $k\in \{2, \ldots, K\}$, and perform a radix sort on
the columns of this matrix (which correspond to the elements $t$ of the permutations). Group together
all elements $t$ that have the same column vector, i.e. the same sign in all permutations. For
each group $s$, the elements of the group that are pushed on $R$   are progressively 
chained together immediately after $\PushLR(\bt,t+1)$ (step 8 in $\LR$-Search), and the pointer
$\Fir(s)$ to the first element of the chain is updated as needed. Then,
for each $t$, $\Rt(t)^*$ is  defined as $\Fir(s)$, where $s$ is the group of $t$. All these operations are done in $O(Kn)$ time and $O(n)$ additional 
space, assuming that radix sort does not really create the indicated matrix, but rather 
manipulates column numbers and checks the values directly on the vectors $\sign_{P_k}$ (which
are not modified).

{\bf Task 2.}  The relative positions of $t$ and $t+1$ in each $P_k$ are checked in $O(1)$  
using the functions $\Pm_k()$, and thus $\Position(t,t+1)$ is computed in $O(K)$ time and
$O(1)$ additional space, for each $t\in\n$.
\bigskip

With this supplementary information, the \out\, procedure is presented in Algorithm \ref{algo:filterconserved}.

\bex
In Figure \ref{fig:LRSearchcv}, conserved intervals are computed for $\PK=\{P_1,P_2\}$,
where $P_1=\Id_7$ and $P_2=(1\, \mbox{-3}\, \mbox{-2}\,\, 6\, \mbox{-4}\, \mbox{-5}\,\, 7)$. Note, for instance, that 
$m_6=4$ but $b_6=3$, by the definition of $\bt$,
indicating that $4$ cannot be the delimiter of a conserved interval containing 6 and 7.
Similar remarks are valid for $b_5, B_3, b_3$ and $B_1$. During the execution of $\LR$-Search,
\out\, does not output $(4..5)$ since $Position(4,5)=false$, but outputs $(2..3)$ and
 $(1..7)$ for which variable $Position()$ is true and the signs are compatible. These are the conserved intervals of $\PK$.
\eex

\begin{thm}
Algorithm $\LR$-Search with settings $\BL, \BR$ and \out$\,$ above solves Conserved-\ISP\, for $K$ 
signed permutations in $O(Kn+N)$ time and $O(n)$ additional space, where $N$ 
is the number of conserved intervals of $\PK$.
\label{thm:cv}
\end{thm}

We start by showing that:

\begin{fait}
An interval $(t..x)$ with the properties {\rm (1), (2)} in Remark \ref{rmk:charcons} also 
satisfies property {\rm (4)} iff $Position(t,t+1)$ is true.
\label{fait:pos}
\end{fait}

\noindent{\bf Proof of Claim \ref{fait:pos}.}   The "$\Rightarrow$" part is obviously true.
For the  "$\Leftarrow$" part, assume by contradiction and without any loss of generality
that $k$ and $x$, with $x>t$, exist such that $t$ is positive and $\Pm_k(x)<\Pm_k(t)$. Then,
the hypothesis that $Position(t,t+1)$ is true insures that $\Pm_k(t)<\Pm_k(t+1)$,
thus $t+1$ does not belong to the interval delimited by $t$ and $x$. Consequently,
condition (1) in Remark \ref{rmk:charcons} is contradicted. The reasoning is similar if $t$
is negative.$\Box$
\bigskip

\noindent{\bf Proof of Theorem \ref{thm:cv}.}  
We show that $(t..x)$ is output by $\LR$-Search  with the given parameters iff it is
a conserved interval. Recall that, by Remark \ref{rmk:cond}, $\Ra(t)$ in 
Algorithm \ref{algo:filtercommon} is the same as $\Ra^t(t)$. We
assume without loss of generality that $t$ is positive.

{\sf Proof of "$\Rightarrow$:"} According to Theorem \ref{thm:B}, the set of intervals 
computed by the algorithm $\LR$-Search is the set of common intervals
(ignoring the signs) $(t..x)$ of $\PK$ with $t=\bt=\min\{b_w\,|\, t\leq w\leq x-1\}$ and 
$x=B_{x-1}=\max\{B_w\,|\, t\leq w\leq x-1\}$.
Then, for each output $(t..x)$ and each $k$,  the corresponding interval on $P_k$ 
has property (1) in Remark \ref{rmk:charcons}.  To check property (2), notice that by Theorem \ref{thm:B}
we have $t=\bt$ and $x=B_{x-1}$, thus according to the conditions on $\BL()$ and $\BR()$: 

\begin{equation*}
\mint\leq t=\bt=\BL(t)\leq\mint\,\,  \mbox{\rm and}\,\, M_{x-1}\geq x=B_{x-1}=B(x-1)\geq M_{x-1}
\end{equation*}

\noindent We deduce that $\bt=\mint=t$ and $B_{x-1}=M_{x-1}=x$. By Theorem \ref{thm:B},
we know that $\bt\leq b_w$ and $B_{x-1}\geq B_w$ for all $w$, $t\leq w\leq x-1$. Assume 
by contradiction that $t$ is not a delimiter of the interval of $P_k$ made of $t, t+1, \ldots,
x$.  Then, there is some $w$, 
$t+1\leq w\leq x-1$ such that $t$ is between $w$ and $w+1$ on $P_k$. Then $m_w\leq m^k_w=t<w$ and 
thus $b_w=\BL(w)\leq m_w-1\leq t-1$. But then $\bt=t>t-1\geq b_w$,
a contradiction. The reasoning is similar for $x$. Property (2) is proved.

Property (3) is insured by the interpretation of $\Rt(t)^*$ and $\Rb(t)$, as well as
by steps 1-3 and 7 in \out. Claim \ref{fait:pos} and step 4 in \out\, guarantee that
the property (4) holds. Thus $(t..x)$ is a conserved interval.

{\sf Proof of "$\Leftarrow$:"}
We have  $\bt=\BL(t)=\mint=t$ and $B_{x-1}=\BR(x-1)=M_{x-1}=x$, by the definition of $\BL, \BR$ and since a conserved interval is a common interval. 
By Theorem \ref{thm:B}, we deduce that $x\in\Ra^t(t)$.   
Furthermore, we use properties (1)-(4) in Remark \ref{rmk:charcons}  and show that no 
interval with these properties is 
forgot by \out. By contradiction, if this was the case, then $x$ would be eliminated by 
the condition in step 4 of \out. But then, by Claim
\ref{fait:pos} the interval $(t..x)$ cannot satisfy property (4) in Remark \ref{rmk:charcons}.
This contradicts the assumption that $(t..x)$ is conserved.

It is easy to see that $\bt$ and $\Bt$ may be computed in $O(Kn)$ time and $O(n)$ additional
space using the values $\mint^k,\maxt^k$ (and avoiding to store all these values), thus we may apply Theorem \ref{thm:complex}.
 \out$\,\,$clearly has running time proportional to the number of output intervals, since
the {\bf while} loop in step 5 is executed only when the condition in step 4 is true.
Moreover, the {\bf while} loop has running time proportional to the number
of output intervals. Theorem \ref{thm:complex} finishes the proof.$\Box$
\bigskip

\begin{figure}[t]
\vspace*{-1cm}
\begin{center}
\hspace*{-1cm}\includegraphics[width=17cm]{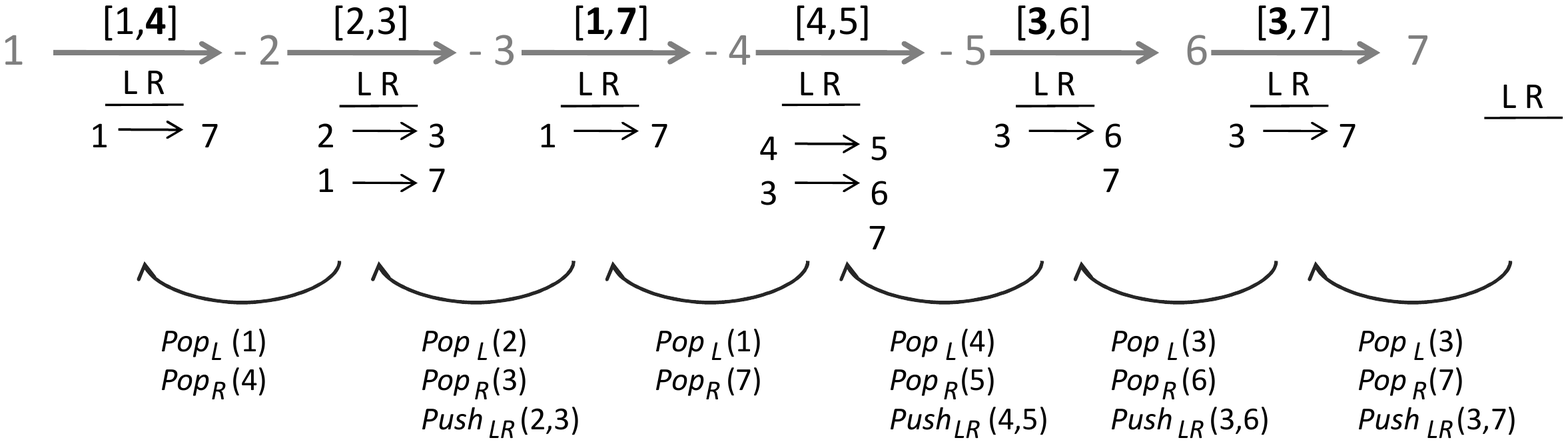}
\end{center}
\vspace*{-6.5cm}
\caption{The $\LR$-Search algorithm for conserved intervals when $P_2=(1\, \mbox{-3}\, \mbox{-2}\,\, 6\ \mbox{-4}\, \mbox{-5}\,\, 7)$.
The chains in $R$  may be easily deduced using the signs. Bounds $\bt$ and $\Bt$ are in bold font whenever they
are not equal to $\mint$ and respectively $\maxt$. }
\label{fig:LRSearchcv}
\end{figure}

\section{Finding subclasses of common, nested and conserved intervals}\label{sect:next}

\subsection{Irreducible common intervals, and same-sign common intervals}\label{subsect:irrcommon}

Irreducible common intervals have been defined in  \cite{heber2011common} as follows. Let $G$ be the graph whose vertices are all the common intervals 
of $\PK$, and whose edges are the pairs of non-disjoint common intervals. Then
a common interval $I$ is {\it reducible} if it is the set union of some of its proper sub-intervals
that are common intervals and induce together a connected subgraph of $G$. Otherwise, $I$ is {\it irreducible}.

Consider now the total order on the set of common intervals of $\PK$ given by $(t_1..x_1)< (t_2..x_2)$
iff either $t_1>t_2$, or $t_1=t_2$ and $x_1<x_2$. For each $w$ with $1\leq w\leq n-1$, let $\Inf(w)$
denote the smallest, with respect to this order, common interval of $\PK$  containing $w$ and $w+1$.  It is shown in 
\cite{heber2011common} that:

\begin{fait}\cite{heber2011common}
The set of irreducible intervals of $\PK$ is the set  $\{\Inf(w)\,|\, 1\leq w\leq n-1 \}$. 

\label{fait:I}
\end{fait}

Then \cite{heber2011common} proposes an algorithm to solve IrreducibleCommon-\ISP\, in linear time.
Another linear time algorithm is obtained by appropriately filtering the results of our $\LR$-Search
algorithm, as shown below.

\begin{fait}
Let $w$ be an integer such that $1\leq w\leq n-1$. The interval $\Inf(w)$ is the common interval $(t..x)$ of
$\PK$ such that  the couple $(n-t,x)$ of integers is minimum with respect to the lexicographic order with 
the property $t\leq w<x$.
\label{fait:irr}
\end{fait}

\noindent{\bf Proof of Claim \ref{fait:irr}.} The condition involving the lexicographic order means
that we first maximize $t$, and then minimize $x$. By contradiction, assume
that the interval $(t..x)$ defined in this way is not $\Inf(w)$. 
Now, $\Inf(w)$ is minimum according to the total order. We must then have 
either $t<\mini(\Inf(w))$ (the smallest
element in $\Inf(w)$), or $t=\mini(\Inf(w))$ and $x>\maxi(\Inf(w))$. But this contradicts 
the choice of $t$ and $x$.$\Box$
\medskip

The $\LR$-Search algorithm we propose here uses the same $\M$-profile for $\PK$ as common intervals, that is
$\BL(i) :=m_i$ and $\BR(i) :=M_i$. However, the \out\, algorithm is in this case a refinement of the 
initial \out\, procedure in Algorithm \ref{algo:filtercommon}.
Algorithm \ref{algo:filterirr} uses   a stack $S$ to store the values
$w$ for which $\Inf(w)$ has not been found yet, in increasing order from top to bottom of $S$. For a given $w$,
the first interval $(t..x)$ such that $t\leq w<x$ found by $\LR$-Search is $\Inf(w)$, by Claim \ref{fait:irr}. 
However, such a
value $x$ may be located in $\Ra^t(t)$ after a sequence of useless values $x'$. At each moment
of the execution of \out, say that  a value $x$ on $R$ is {\it untrusty} if at least one interval 
$(t'..x)$ has been already  output  (meaning that $x$ has possibly become useless), and {\it trusty} otherwise. 
Call a {\it strip} any maximal sequence of untrusty consecutive elements on $R$ that are also consecutive integers.

In order to insure  the best running time for our algorithm, we consider that each element $x$ on $R$ but the bottom
of $R$ has, in addition to its successor $\next(x)$ on $R$, another successor denoted $\nextu(x)$:

\begin{enumerate}
\item[$\bullet$] if $x$ is trusty, then its successor $\nextu(x)$ is $\next(x)$;
\item[$\bullet$] if $x$ is untrusty and is the head (i.e. the first)  element in its strip, then $\nextu(x)$ is the
first element following the strip (if such an element exists). Note that this element may be either trusty or
untrusty.
\item[$\bullet$] if $x$ is untrusty and it is not the head element in its strip, then $\nextu(x)$ is by definition identical to $\nextu(x')$, where $x'$ is the head element in the strip of $x$. Then, by definition, $\nextu(x)$ {\it changes} iff
the head of its strip changes.
\end{enumerate}

\br
Note that the state (trusty or untrusty) of an element may be easily computed in \out. Moreover, 
the successor $\nextu(x)$ of an element $x$ is computed in $O(1)$ when $x$ is pushed on $R$ by
$\PushLR$. Furthermore, when an element $x$ becomes untrusty (step 10 in \out), the successors $\nextu()$ 
change in the strip before and possibly in the strip after $x$ on $R$ (in the latter case, the head of the strip may change, thus
 - as indicated before - the successors $\nextu()$ also change). We call Update$(\nextu())$ 
the procedure performing this update in Algorithm \ref{algo:filterirr} (step 10). We do not give its details here,
but discuss it in the proof of Theorem \ref{cor:irr}.
\label{rmk:nextu}
\er

\begin{algorithm}[t,boxed]
\caption{The \out$\,\,$algorithm for irreducible common intervals} 
\begin{algorithmic}[1]
\REQUIRE Pointers $\Rt(t),\Rb(t)$ to the first and last element of $\Ra(t)$ (possibly equal to $nil$)\\ 
Stack $S$ and values $\nextu()$ output by the preceding  \out\, call (except when $t=n-1$).

\ENSURE  All irreducible common intervals $(t..x)$ of $\PK$ with fixed $t$.

\IF{$t=n-1$}
\STATE Let $S$ be an empty stack
\ENDIF
\STATE Push $t$ on $S$ {\sl \hfill // $\Inf(t)$ has not been found yet}
\IF {$\Rt(t)\neq nil$}
\STATE $x^{\top}\leftarrow $ the target of $\Rt(t)$; $x^{\bot}\leftarrow $ the target of $\Rb(t)$
\STATE $x\leftarrow x^{\top}$
\WHILE{$x\leq x^{\bot}$ and $S$ is not empty and ($\tp(S)<x$ or $x$ is untrusty)}
\IF{$\tp(S)< x$}
\STATE Output the interval $(t..x)$; Update($\nextu()$){\sl \hfill //$x$ becomes untrusty}
\WHILE{ $S$ is not empty and $\tp(S)<x$}
\STATE Pop $\tp(S)$ from $S$ {\sl \hfill // for $\tp(S)$, interval $\Inf(\tp(S))$ has been output}
\ENDWHILE
\STATE $x\leftarrow \next(x)$ \hfill {\sl //or $n+1$ if $\next(x)$ does not exist}
\ELSE
\STATE $x\leftarrow \nextu(x)$\hfill {\sl //or $n+1$ if $\nextu(x)$ does not exist}
\ENDIF
\ENDWHILE
\ENDIF 

\end{algorithmic}
\label{algo:filterirr}
\end{algorithm}

\bex
On the example in Figure \ref{fig:LRSearch}, the $\LR$-Search algorithm with the \out\, procedure in Algorithm
\ref{algo:filterirr} first outputs (when $t=4$) $(4..5)$ and $(4..6)$ (which are $\Inf(4)$ and $\Inf(5)$). The
values $5$ and $6$ become untrusty (step 10). When $t=3$, the interval $(3..6)$ is output, which is
$\Inf(3)$. Finally, with $t=1$, intervals $(1..2)$ and $(1..3)$ are successively output (they are $\Inf(1)$ and
$\Inf(2)$). The stack $S$ still contains $6$ and the next value in $\Ra(1)$ is $x=6$. Thus $\tp(S)=x=6$ in  
step 8, and $x$ is untrusty. In step 16, $x$ is updated to the value $\nextu(6)$ which is 7.
The interval $(1..7)$, which is $\Inf(6)$, is output.
\eex

\begin{thm}
Algorithm $\LR$-Search with settings $\BL, \BR$ as for common intervals and the more restrictive \out\, procedure in Algorithm \ref{algo:filterirr} solves IrreducibleCommon-\ISP\ for $K$ 
permutations in $O(Kn)$ time and $O(n)$ additional space.
\label{cor:irr}
\end{thm}

\noindent{\bf Proof of Theorem  \ref{cor:irr}.} The decreasing order of $t$ in the {\bf for} loop of the $\LR$-Search
algorithm and the increasing values of the elements in $\Ra^t(t)$ show that the
intervals $(t..x)$ are considered according to the lexicographic order required by 
Claim \ref{fait:irr}. Values $w$ are pushed on $S$ as soon as $w$ is considered
(step 4 in \out) and are discarded from $S$ iff the interval with maximum $t$ and
minimum $x$ containing $w$ and $w+1$ is output (steps 10 and 12).   By Claim \ref{fait:irr}, 
this interval  is $\Inf(w)$. Thus, all the intervals output by the algorithm are irreducible.

Conversely, assume by contradiction that some interval $\Inf(w)$ is not output by the algorithm, and let
$w$ be the smallest such value. Let $\Inf(w)=(t..x^*)$ and consider the execution of \out\, for $t$. 
Since  $(t..x^*)$ is not output by the algorithm, 
the execution of the {\bf while} loop in step 8 satisfies one of the following conditions:

(i) either it misses $x^*$ by skipping it in step 16, 

(ii) or stops before $x^*$ is reached. 

\noindent Notice that $x^*\in\Ra^t(t)$ (by Theorem~\ref{thm:B}).
Let $x'$ be the largest element smallest than $x^*$  for which the condition in step 8 of \out\, is tested,
and consider the state of the stack $S$ immediately after this test.

Let us show that $\tp(S)=w$. Note that the elements in $S$ are in increasing order from top
to bottom. By contradiction, if we assume $u :=\tp(S)<w$ then
the minimality of  $w$ insures that $\Inf(u)$ is output by the algorithm $\LR$-Search with
the given settings. Then $\Inf(u)=(t_0..x_0)$ with $t_0\leq t$, since $u$ is still on $S$.
Moreover, $u>t$ since $\PushLR$ has pushed on $R$ only elements $t'+1$ with $t'\geq t$.
Now, we cannot have $t_0<t$ since then $t_0<t< u < w<x^*$ and thus $(t..x^*)$ is smaller
than $\Inf(u)$ and contains both $u$ and $u+1$, a contradiction. We thus have
$t_0=t$. Now, $(t..x_0)$ is certainly output by \out, and this has not been done yet when $x'$ is
considered (otherwise, $u$ would have been discarded from $S$). We deduce that $x'$ 
does not stop the execution of the {\bf while} loop (i.e. case (i) before does not hold). Thus 
$u<x'$ or $x'$ is untrusty in step 8, and we are in case (ii).
The former case ($u<x'$) would contradict the maximality of $x'$, because then the condition in
step 9 is true and thus step 14 performs $x\leftarrow next(x')$,  which brings into step 8 a value 
larger than $x'$. The latter case ($x'$ is untrusty)  implies that the next 
trusty value is larger than $x^*$ (since $x^*$
is skipped because of $x'\leftarrow \nextu(x')$), and thus $x_0>x^*$ and $(t..x^*)<\Inf(u)$
contradicts the minimality of $\Inf(u)$. Thus $\tp(S)=w$.
\medskip

Consider now the cases (i) and (ii) before, and let us show that they both lead to a contradiction (thus
proving our theorem).

\begin{enumerate}
\item[(i)] The execution of the {\bf while} loop in step 8 misses $x^*$ by skipping it in step 16.

In this case, $x'$ is the value for which step 16 has been executed.
Then $x'$ and $x^*$ are untrusty (by the definition of $\nextu()$), $x^*>x'$ (otherwise $x^*$ is not skipped) 
and the values in $\Ra^t(t)$ between $x'$ and $x^*$ are consecutive and untrusty (they belong to the
same strip). Moreover, step 16 has 
been executed, so that the condition in step 9 of \out\, is not verified. Thus $\tp(S)\geq x'$. 
Since $\tp(S)=w$ and $w<x^*$ (recall that $(t..x^*)$ contains $w$ and $w+1$) we deduce that $x^*>w\geq x'$.
Then $w$ is between $x'$ and $x^*-1$ on the strip, and $w+1$ is between $x'+1$ and $x^*$ on the
strip. Consecutively, $w+1$ is untrusty, thus an interval $(t'..w+1)$ with $t'>t$ has been
already output by \out. But this interval is smaller that $(t..x^*)$ and should therefore be
$\Inf(w)$, a contradiction. 
\item[(ii)] The execution of the {\bf while} loop in step 8 stops before $x^*$ is reached. 

In this case, $x'$ satisfies $x'<x^*$ and $x'\leq \tp(S)$ and
$x'$ is trusty. Now, $x'-1$ is not on $S$, since $w=\tp(S)$ and elements in $S$ are in increasing
order from top to bottom. Thus, by the minimality of $w$, $\Inf(x'-1)$ is output by the algorithm.
Let $\Inf(x'-1)=(t''..x'')$ and notice that $x''> x'$ ($x'$ must be contained in $\Inf(x'-1)$ but
$x'$ has not become untrusty) 
and $t''>t$ (otherwise, $x'-1$ should be on $S$). Then $(t..x')\cap(t''..x'')=(t''..x')$ and it is 
a common interval containing both $x'-1$ and $x'$ and which is smaller than $\Inf(x'-1)$, a contradiction.
\end{enumerate}

The correctness of the algorithm is proved. We discuss now the running time of the algorithm.
Leaving apart temporarily the update of $\nextu()$, the running time of \out\, is proportional 
to $v_t$, where $v_t$ is the  number of elements $w$ for which $\Inf(w)$ has smallest value $t$ (these
elements are discarded in step 12).  Over all the executions
of \out, we obtain $O({\it v})$, where ${\it v}=\Sigma_{1\leq t<n}v_t$.
Now, ${\it v}$ is in $O(n)$,  since it counts the total number of elements $w$.

We need now to show that the update of $\nextu()$ in step 10 of \out\ may be done in $O(1)$.
For this, we see each strip $s$ as a set $T(s)$ for which: (a) its minimum element is also its
representative element, denoted $r_{T(s)}$, such that, by definition, $\nextu(r_{T(S)})$ gives the successor 
$\nextu()$ of all the elements in the strip $s$, and (b) its maximum element is denoted $m_{T(s)}$, such that
the set of elements in $T(s)$ is $\{r_{T(s)}, r_{T(s)}+1, \ldots, m_{T(s)}\}$. 

Then it is sufficient to update $\nextu(r_{T(S)})$  in order to update $\nextu()$ for all the elements in $s$. 
Now, notice that strips may be changed in two ways: 

\begin{enumerate}
\item[(1)] the instruction $\PopR$ in $\LR$-Search may perform
deletions from some strip $s$. However, we keep the deleted elements in 
the set $T(s)$ (we only mark them as deleted) and thus the representative $r_{T(s)}$, as well as the set $T(s)$, are 
unchanged by these  deletions (although the strip itself is reduced).
\item[(2)] in step 10 of \out, the element $x$ becomes untrusty and then the strip
immediately preceding $x$ (if any), the strip formed by $x$ alone, and the strip immediately
following $x$ (if any) may concatenate (altogether or only two of them, which have consecutive elements).
Concatenations of strips imply unions of the corresponding sets, and thus changes of the representative elements.
\end{enumerate}

In conclusion, updating the strings and being able to find the representative of each of them
places us in the context of a Union-Find structure.  The sets are the sets $T(s)$, containing
the elements of $s$ as well as all the elements previously in $s$ and discarded from $R$ by $\PopR$. 
The unions between sets are given by the concatenations between strips, whereas the find 
operation for an element $y$ seeks the representative element of the strip containing $y$.
Again, we are in the particular case where unions always involve sets of consecutive 
integers (this is the case when two strips are concatenated). Thus, according
to the result in \cite{itai2006linear}, the implementation of unions and finds may be done in
time linear with respect to the number of union and find operations.

Consequently, the operation Update$(\nextu())$ in step 10 needs to concatenate if necessary
two or three of the strips indicated in (2) and to update $\nextu()$ for their 
representative elements. Over all the executions of \out, these operations are done in $O(n)$ 
and the running time of
the algorithm is proved.$\Box$
\bigskip

\noindent{\sl Same-sign common intervals}

In \cite{heberclusters}, an algorithm is proposed for finding, in $K$ signed 
permutations, all common intervals whose elements have the same sign inside each
permutation. To solve this case, our $\LR$-Search algorithm needs to 
(1) preprocess $\PK$ to compute, for each $t$ and $k$, the minimum $x_k^t>t$, such 
that $t$ and $x_k^t$ have different signs in $P_k$; and (2) stop to output intervals 
in the \out\, procedure (step 4 in Algorithm \ref{algo:filtercommon}) as soon as
the first $x$ with $x\geq \min\{x_k^t\, |\, 2\leq k\leq K\}$ is found. 

For a fixed $k$, task (1) is easily done in $O(n)$ by considering the elements of $P_k$
in decreasing order, and remembering the sign changes. The global time required
for this task is thus in $O(Kn)$, whereas the additional space may be limited to
$O(n)$ by computing the $\min()$ values above progressively.

\subsection{Maximal nested intervals}\label{subsect:maxnested}

In \cite{hoberman2005incompatible}, authors define a 
nested interval $I$ to be {\it maximal} if it is not included in a nested 
interval of size $|I|+1$. In \cite{blin2010finding}, an efficient algorithm is proposed to solve
MaximalNested-\ISP\, for $K=2$. Not surprisingly, $\LR$-Search 
works in this case too, with an appropriate filtering algorithm.

To this end, note that:

\begin{fait}
A nested interval $(t..x)$ is maximal iff it satisfies the two following conditions:
\begin{enumerate}
\item[$(a)$] $(x,x+1)$ is a gap with $x\in V_t$, or $x=y_t$
\item[$(b)$] $b_{t-1}<t-1$ or $B_{t-1}>x$.
\end{enumerate}
\label{fait:mnest}
\end{fait}

\noindent{\bf Proof of Claim \ref{fait:mnest}.} By definition, $(t..x)$ is maximal iff it
is nested but neither $(t..x+1)$ nor $(t-1..x)$ are nested. By Claim \ref{fait:nest}, 
this is equivalent to $x\in V_t$, $x+1\not\in V_t$ and $x\not\in V_{t-1}$ (or $V_{t-1}$ is
not defined).
Now, $x\in V_t$ and $x+1\not\in V_t$ simultaneously hold iff property $(a)$ in the claim is satisfied.

 Moreover, $x\in V_t$ and $x\not\in V_{t-1}$ (or $V_{t-1}$ is not defined)
iff, at the end of step 9 of {\bf for}$_{t-1}$,  either $t-1$ is not on top of $L$  
or $x$ has been removed from $R$. Equivalently, $t-1> b_{t-1}$
or $x<B_{t-1}$. The claim is proved.$\Box$

\bex
It is easy to see that in Figure \ref{fig:LRSearch}, only intervals $(3..6)$ and $(1..3)$
satisfy these conditions. They are indeed the only maximal nested intervals of  $\PK$.
\eex

Then we have:

\begin{thm}
Algorithm $\LR$-Search with settings $\BL, \BR$ as for nested intervals and an appropriate \out\, algorithm  solves MaximalNested-\ISP\, for $K$ 
permutations in $O(Kn+N)$ time and $O(n)$ additional space, where $N$  is the number of maximal nested intervals of $\PK$.
\label{cor:nested}
\end{thm}

\noindent{\bf Proof of Theorem \ref{cor:nested}.} According to Claim \ref{fait:mnest}, \out\, should output only the intervals that
satisfy conditions $(a)$ and $(b)$. The latter condition is easy to test.
The former one needs to find each gap, as well as $y_t$, in $O(1)$.  For this, it
is sufficient to compute and store, for each $r\in R$,  the pointer 
$R^g(r)$ defined in Remark \ref{rem:W}. Values $R^g(t+1)$ must be initialized  
immediately after $\PushLR(b_{t},t+1)$, in step \fort\, of $\LR$-Search.  They do not need to be updated.

Then it is sufficient to modify \out\,\, in Algorithm \ref{algo:filternested} by replacing steps  10-15 with steps 
10-15 in Algorithm \ref{algo:filternested2}.

\begin{algorithm}[t,boxed]
\caption{The \out$\,\,$algorithm for maximal nested intervals}
\begin{algorithmic}[1]
\setalglineno{10}
\STATE $x\leftarrow R^g(x^{\top})$; Compute $y_t$
\WHILE{$x\leq y_t$ and ($b_{t-1}<t-1$ or $B_{t-1}>x$)}
\STATE Output the interval $(t..x)$
\STATE $W[t-1]\leftarrow x$ {\sl \hfill // $t$ passes down to $t-1$ its largest $x$ such that $(t..x)$ is output}
\STATE $x\leftarrow R^g(\next(x))$ {\sl \hfill //or $x\leftarrow n+1$ if $\next(x)$ does not exist}
\ENDWHILE

\end{algorithmic}
\label{algo:filternested2}
\end{algorithm}

Computing $y_t$ in $O(1)$ is possible according to Remark \ref{rem:W}. The other modifications 
aim at precisely selecting the elements with properties $(a)$ and $(b)$ in Claim \ref{fait:mnest}.
Notice that when a first value $x$ not satisfying properties $(a)$ and $(b)$ is found in step 11, 
it is clear that no other subsequent value  $x'$ (necessarily $x'>x$) will satisfy properties $(a)$ and $(b)$.
The resulting \out\, procedure then finds all the maximal nested intervals, by Claim
\ref{fait:mnest}, runs in global time proportional to the number of output intervals,
and globally uses $O(n)$ additional space.$\Box$ 
  
\subsection{Irreducible conserved intervals}\label{subsect:irrcons}

In \cite{BergeronSim}, authors define a conserved interval to be {\em irreducible} if it is not the union of smaller
conserved intervals. They also give an efficient algorithm to solve IrreducibleConserved-\ISP. Such an algorithm may also be obtained using $\LR$-Search and
the following easy result:

\begin{fait}
Let $(t..x)$ be a conserved interval of $\PK$. Then $(t..x)$ is irreducible 
iff  

\begin{equation*}
x=\min\{h \, |\, t<h\, \mbox{\rm and}\, (t..h)\, \mbox{\rm is a conserved interval of}\, \PK\}.
\end{equation*}
\label{fait:irrcons}
\end{fait}

\noindent{\bf Proof of Claim \ref{fait:irrcons}.} In \cite{BergeronSim} it is shown that two different
irreducible intervals are either disjoint, or nested with different endpoints or else
overlapping on one element. The conclusion follows.$\Box$
\medskip

We deduce:

\begin{thm}
Algorithm $\LR$-Search, with settings $\BL, \BR$ as for conserved intervals and a simplified \out$\,$ procedure,
solves IrreducibleConserved-\ISP\, for $K$ permutations in $O(Kn)$ time and $O(n)$ additional space.
\label{cor:irrcv}
\end{thm}

\noindent{\bf Proof of Theorem \ref{cor:irrcv}.} By Claim \ref{fait:irrcons}, it is sufficient to replace the {\bf while} loop in the \out\, procedure with
an instruction that outputs $(t..x^{\top})$.$\Box$

\section{Conclusion}\label{sect:concl}

The $\LR$-stack we introduced in this paper is a simple data structure, of which
we noted at least two advantages: it is powerful (we had two applications of it in this
paper), and it is algorithmically efficient, since it makes use of the 
efficiency reached by the Union-Find-Delete algorithms. 

Using $\LR$-stacks, our algorithmic framework $\LR$-Search succeeds in proposing a
unique approach for dealing with common intervals and their subclasses, for an arbitrary number
$K$ of permutations. The computation
of the interval candidates is driven by the $\M$-profile and the bounding functions, whose role is
to guarantee that all interval candidates satisfy the content-related constraints.
Afterwards, the \out\, procedure chooses between the candidates those that satisfy
the supplementary constraints defining a precise subclass.

All the algorithms resulting from this approach are as efficient as possible. They
allowed us to prove the power and the flexibility of our approach. Among them,
the algorithms searching for nested and maximal nested intervals of $K$ permutations, 
with $K>2$, solve previously unsolved problems. 

\bibliographystyle{plain}
\bibliography{biblio}
\end{document}